\documentclass[12pt]{article}
\usepackage{amssymb,amsmath, epsfig}
\include{epsf}
\usepackage{bbm}
\usepackage{dsfont}
\allowdisplaybreaks

\usepackage{subfig}
\usepackage{mathtools}
\usepackage{graphicx}

\usepackage{hyperref}
\usepackage{color}

\usepackage{tikz,pbox}
\usetikzlibrary{shapes,arrows}
\usetikzlibrary{calc,decorations.markings}

\newcommand{\mytikz}[1]{
	\pbox{\textwidth}{\begin{tikzpicture}[>=stealth,decoration={
    markings,
    mark=at position 0.5 with {\arrow{>}}}]	
		#1
	\end{tikzpicture}}
}

\usepackage{amsfonts}
\usepackage{amscd}
\usepackage[all]{xy}
\advance \textheight by \topskip
\makeatletter

\def\ket#1{\left| #1\right\rangle}

\newcommand{\be}{\begin{equation}}
\newcommand{\ee}{\end{equation}}
\newcommand{\beq}{\begin{equation}}
\newcommand{\eeq}{\end{equation}}
\newcommand{\bea}{\begin{eqnarray}}
\newcommand{\eea}{\end{eqnarray}}
\def\beqa{\begin{eqnarray}}
\def\eeqa{\end{eqnarray}}
\def\nn{\nonumber}

\newcommand{\eq}{\begin{equation}}
\newcommand{\eqa}{\begin{eqnarray}}
\newcommand{\en}{\end{equation}}
\newcommand{\ena}{\end{eqnarray}}

\newcommand{\ses}{ {\setminus} } 

\newcommand{\cN}{ {\mathcal N} } 

\newcommand{\IC}{{\mathbb C}}

 \newcommand{\cK}{\mathcal{K}}

\newcommand{\cZ}{ \mathcal{Z} }

\newcommand{\diag}{ {\rm Diag} } 
\newcommand{\Sym}{ {\rm Sym} } 
\newcommand{\plog}{ {\rm Plog} }

\newcommand{\bdel}{ {\boldsymbol{\delta}} }

\newcommand{\la}{\langle}
\newcommand{\ra}{\rangle}

\newcommand{\mC}{\mathbb{C}} 

\newcommand{\Tr}{{\rm Tr}}
\newcommand{\tr}{{\rm tr}} 
\newcommand{\cy}{{\bf{c}}}
\newcommand{\Dim}{ {\rm Dim } } 
\def\s{ \sigma }
\def\g { \gamma }
\def\ws{ \widetilde\sigma }
\def\wt{ \widetilde\tau }

\newtheorem{theorem}{Theorem}[section]

\newtheorem{lemma}[theorem]{Lemma}

\newtheorem{proposition}[theorem]{Proposition}

\newcommand{\proof}{\noindent{\bf Proof.}\;} 
\newcommand{\qed}{{\hfill $\Box$}}

\def\Qun#1{Q^{R#1_1,R#1_2,R#1_3,\tau#1}}
\def\cbbb#1#2{\sum_{q_l} C^{R#1_1,R#1_2;R#1_3,\tau#2}_{q_1,q_2;q_3} 
B^{R#1_1;\,tr}_{q_1} 
B^{R#1_2;\,tr}_{q_2} 
B^{R#1_3;\,tr}_{q_3}}
\def\kun#1#2#3#4#5{\sum_{#1_i \in S_n[S_2]} \sum_{#5 \in S_{2n}} 
#1_1 #2 #5 \otimes  #1_2 #3 #5 \otimes  #1_3 #4 #5}

\begin{document}
\title{
{\bf On the counting of $O(N)$ tensor invariants} }
\author{
{\sf   Remi C. Avohou${}^{a,b,c}$\thanks{e-mail: avohou.r.cocou@mail.huji.ac.il}}
{\sf   Joseph Ben Geloun${}^{d,c}$\thanks{e-mail: bengeloun@lipn.univ-paris13.fr}},   
{\sf Nicolas Dub${}^{d}$\thanks{e-mail: dub@lipn.univ-paris13.fr}}, \\
{\small ${}^{a}${\it Einstein Institute of Mathematics, The Hebrew University of Jerusalem, }} \\
{\small {\it Giv’at Ram, Jerusalem, 91904, Israel }} \\
{\small ${}^{b}${\it Ecole Normale Sup\'erieure, B.P 72, Natitingou, Benin  }} \\
{\small ${}^{d}${\it LIPN, UMR CNRS 7030, Institut Galil\'ee, Universit\'e Paris 13, 
Sorbonne Paris Cit\'e, }} \\
{\small {\it  99, avenue Jean-Baptiste Cl\'ement, 93430 Villetaneuse, France}} \\
{\small ${}^{c}${\it  International Chair in Mathematical Physics and Applications}} \\
{\small {\it ICMPA-UNESCO Chair, 072Bp50, Cotonou, Benin}} \\
}
\date{\today}

\maketitle

\vskip-1.5cm

\vspace{2truecm}

\begin{abstract}\noindent
$O(N)$ invariants are the observables of real tensor models.  We use regular colored graphs to represent these invariants, the valence of the vertices of the graphs relates to the  tensor rank.  We enumerate $O(N)$ invariants as $d$-regular graphs, using permutation group techniques. We also list their generating functions and give (software) algorithms computing their number at an arbitrary rank and an arbitrary number of vertices. As an interesting property, we reveal that the algebraic structure which organizes these invariants differs from that of the unitary invariants. 
The underlying topological field theory formulation of the rank $d$ counting shows that it corresponds to counting of coverings of the $d-1$ cylinders sharing the same boundary circle and  with $d$ defects.
At fixed rank and fixed number of vertices, an associative semi-simple algebra with dimension the number of invariants naturally emerges from the formulation.  
Using the representation theory of the symmetric group, we enlighten a few crucial facts: the enumeration of $O(N)$ invariants gives a sum of constrained Kronecker coefficients;  there is a representation theoretic orthogonal base of the algebra that reflects its dimension; normal ordered 2-pt correlators of the Gaussian models evaluate using permutation group language, and further, via representation theory, these functions provide other representation theoretic orthogonal bases of the algebra. 
\end{abstract}

\newpage 

\tableofcontents

\section{Introduction}
\label{intro}

Since their inception \cite{ambj3dqg,mmgravity, sasa1}, random tensor models offer a framework for studying random discrete geometries as they aim at extending the success of matrix models \cite{Di Francesco:1993nw} 
in describing 2D quantum gravity, to higher dimensions. 
The main goal of this approach is to devise a transition of discrete geometries to continuum geometries
in any dimension. 
It is however only recently that random tensor models have witnessed significant progress \cite{razvanbook} with the advent of a new large $N$ expansion  generalizing `t Hooft genus expansion \cite{'tHooft:1973jz} for higher dimensional pseudo-manifolds. The existence of a large $N$ expansion for tensors \cite{Gur4} naturally unveiled  several analytical results, among which the discovery of their critical behavior (branched polymers \cite{Bonzom:2011zz,Gurau:2013cbh}),
 the universality property of random tensors \cite{Gurau:2011kk},  and the discovery of new families of renormalizable non-local quantum field theories with interesting UV  \cite{BenGeloun:2011rc,Geloun:2013saa, Carrozza:2013mna} and nonperturbative behaviors supporting the discovery of new universality classes 
 for gravity \cite{Eichhorn:2017xhy, Eichhorn:2018ylk,Eichhorn:2018phj}. 
 
 More recently, and quite unexpectedly,  tensor models become the center of new attention in condensed matter
  physics: the dominant contributions of the so-called Sachdev-Ye-Kitaev (SYK) model \cite{SYK,maldastan} in the large mode expansion of the disorder match with the large $N$ expansion  of a quantum mechanical tensor model without disorder \cite{witten}. 
For its deep connections with black hole physics and AdS/CFT correspondence, the SYK model embodies a vibrant topic of research. 
The  conjunction of tensor and SYK models has
incidentally produced a new fast-evolving field on which a growing community is working. 

Several, if not all of these studies, heavily rest on the understanding of the combinatorics
  of Feynman graphs and observables of tensor models. To that extent, the  investigations of tensor models have produced a wealth of results. We will focus on two particular contributions on tensor model graphs that the present work extends. 
  
In \cite{BenRamg}, the authors worked out the enumeration of the unitary invariants, 
as observables in complex tensor models. One way of comprehending the theory space of 
rank $d$ complex tensor models is to specify its set of observables.
The latter are merely $U(N)^{\otimes d}$ invariants (at time, we simply call them $U(N)$ or complex tensor invariants).  A convenient manner to represent $U(N)$ invariants defines as a canonical mapping to $d$-regular  bipartite colored graphs
\cite{color}. Stated in this way, the inventory of tensor invariants formulates by uniquely using permutation groups.
One should record that these symmetry group techniques 
and its representation theory have been developed during the last
years  \cite{cjr}--\cite{Diaz:2018xzt}. They turned out to be powerful, flexible and versatile enough to address 
diverse enumeration problems and bijections from scalar field theory, matrix models, to gauge  
(QED, 2D and 4D Yang-Mills) and string theories. In physics,  for instance, they brighten the half-BPS sector of $\cN=4$ SYM \cite{cjr}--\cite{doubcos}. Moreover, unforeseen correspondences arise from these studies,
for instance, counting Feynman graphs in $\phi^4$ scalar field theory relates to string theory 
on a cylinder or  listing Feynman graphs of QED relates to the counting of ribbon graphs \cite{FeynCount}. These correspondences emerge from another interface playing a hinge role between enumeration problems: 
via the Burnside lemma, with each enumeration problem using the symmetric group (and its subgroups), we can associate 
a Topological Field Theory on a 2-complex (named TFT$_2$) with gauge group given by the symmetric group (and its subgroups). Such a formulation also unfolds multiple interpretations of the counting formulae
with links with the theory of covering spaces in algebraic and complex geometry
(see references in \cite{FeynCount}).  
  
The reference  \cite{BenRamg} establishes several enumeration formulae 
pertaining to observables of complex tensor models. Using the Burnside lemma, one recasts that the enumeration 
of $U(N)$ invariants  into a partition of a  permutation lattice gauge field theory, a TFT$_2$. It is via this mapping that 
one elucidates that counting unitary invariants corresponds to counting branched covers of the 2-sphere. 
Branched covers are well known objects in algebraic and complex geometry \cite{kwak}, in topological string theory, and in dimension 2,  they correspond to complex maps \cite{Gal}. Thus, there is an underlying geometry inherited 
by tensor models from the TFT$_2$ formulation that still needs to be understood. There is
however a proviso: the counting formulae are valid when the size $N$ of the tensor indices
are larger than the number of tensors convoluted. More generally, one should resort to a more careful 
study \cite{Diaz:2017kub,Diaz:2018xzt}.   

The study of tensor invariants has a follow-up in \cite{BenSamj}. Their equivalent classes 
are viewed as the base elements of a vector space $\cK_d(n)$, a subspace of $\IC(S_n)^{\otimes d}$, the rank $d$ 
group algebra of the symmetric group $S_n$.  $\cK_d(n)$ shows stability under an associative
product, and it is endowed with a non-degenerate pairing. Therefore, at a fixed rank $d$ and fixed number of
vertices $n$, tensor invariants span a semi-simple algebra.  
(Note that, importantly,  other algebraic structures could set up on tensor invariants 
\cite{Itoyama:2017wjb,Itoyama:2018but, Itoyama:2019oab}. The above structure is however
unique, up to isomorphism.)
As a consequence of the Wedderburn-Artin
theorem, any semi-simple algebra decomposes as a sum of irreducible matrix subalgebras. 
The representation theory of the symmetric group sheds more light on the remaining analysis 
as it enables to reach the Wedderburn-Artin matrix decomposition of the algebra of tensor observables: 
the dimension of the algebra is a sum of squares of the Kronecker coefficients (these are multiplicity 
dimensions in the decomposition of a tensor product of representations in irreps; 
Kronecker coefficients are still under active investigation in Combinatorics and Computational Complexity Theory,
see, for instance, \cite{iken1,Blasiak} and more references therein), each square matching exactly the 
 dimension of a matrix subalgebra. 
 The orthogonal bases of the algebra and its matrix subalgebras have been worked out, 
meanwhile the Gaussian 2pt-correlators also provide new representation theoretic orthogonal bases. 
 
In this paper, we consider $O(N)$ tensor models and their observables 
and investigate if they support the same previous enumeration and algebraic analysis. 
 Fleshed out the first time in \cite{CarrooAdri}, such models extended the large $N$ expansion to real tensors.  
The graphs that determine the $O(N)$ invariants keep the edge coloring but are not bipartite. This naturally leads to a class of observables, wider than that of the $U(N)$ tensor models, by including those that are not orientable. 
To enumerate $O(N)$ invariants, we use a standard counting recipe: 
we use tuples of permutations on which act permutation (sub)groups that define equivalence classes. 
We then count the points in the resulting a double coset space. 
The equivalence relation in the present setting is radically different from the $U(N)$ situation 
and requires more work to obtain a valuable counting formula. 
With their generating functions in hand, we provide software  (Mathematica, Sage) codes to achieve the counting of $O(N)$ observables for any tensor rank. 
We emphasize that our results match the seminal work of Read in \cite{Read} that dealt with the enumeration $k$-regular graphs with $2n$ vertices with $k$ edge coloring. However, 
Read's formula was only evaluated for the $k=3$-regular graphs with $2n=2,4,6$ vertices with edges of $3$ different colors. Our code  extends this counting for any $k$ and any $n$.  We produce  integer sequences that are new (un-reported yet) to the  On-Line Encyclopedia of Integer Sequences \cite{oeis}.

Moreover, seeking other correspondences, we address the TFT formulation of our counting
and show that to count $O(N)$ observables amounts to count covers of glued cylinders with defects
(the rank of tensor relates to the number cylinders and defects). 
 After introducing the algebra of $O(N)$ invariants, we show that it is semi-simple, and 
 as such, it admits a Wedderburn-Artin decomposition. An invariant orthogonal base of the algebra transpires
 in our analysis but, it does not yield the decomposition of the algebra in matrix subalgebras. 
We proceed to the representation theoretic formulation of the counting and its consequences. 
As to be distinguished from the $U(N)$ case, the dimension of the algebra 
is a sum of constrained Kronecker coefficients restricted to partitions will
all even length rows. The  representation theoretic tools exhibit a base of the algebra
the dimension of which directly reflects the sums of constrained Kronecker's. 
The Gaussian 2pt and 1pt-correlators also compute in terms of permutation group formulae. 
A corollary of that analysis is that 2pt-functions, in the normal order,  select a representation theoretic orthogonal base of the algebra. In that sense, the Gaussian integration in the representation Fourier space 
performs as a pairing of observables. 

This paper's structure follows. Section \ref{sect:oninv} sets up our notations for real tensor models
and their $O(N)$ invariants. The following section \ref{section:counting} develops the double coset counting
using permutation group formalism. We also discuss therein the TFT formulation of the counting 
and its consequences,  introduce the basics of the representation theory of the symmetric group, 
and re-interpret the counting in that language. 
 Section \ref{sect:doublcoset} discusses the double coset algebra built out of the $O(N)$ invariants
 and lists its properties. Next, section \ref{sect:correl} details the 2pt- and 1pt-correlators of the Gaussian 
 tensor models and their representation theoretic consequences. 
 Section \ref{spn} briefly lists a few remarks on the counting of invariants of the real symplectic group $Sp(2N)$. 
 The counting principle here is similar to that of the $O(N)$ models, but with subtleties that one should pay attention to. 
 Section \ref{concl}  summarizes this work and draws some of its perspectives.  
 Finally, the paper closes with an appendix that divides into two main parts: 
 an appendix that collects  identities of the representation theory of the symmetric group
 that are useful in the text,  and another appendix that details the software codes that generate the sequences of numbers
 of invariants at sundry tensor ranks $d=3,4,...$.

\section{$O(N)$ invariants and real tensor models}
\label{sect:oninv}
We first set up our  notations in this part. 

Consider $d\geq2$ real vector spaces $V_a$, $a=1,\dots, d$, of respective dimensions $N_a$,  and the group action $\otimes_{a=1}^dO(N_a)$ on $\otimes_{a=1}^dV_a$. Let $T$ be a tensor of rank $d$ with components $T_{i_1, \cdots, i_d}$ transforming under the  tensor product of $d$ fundamental representations of the  groups $O(N_a)$. Each group $O(N_a)$ acts  independently on a tensor index $i_a$ and we can write 
\bea
T_{i_1, \cdots, i_d}^O=\sum_{j_1, \cdots, j_d} O^{(1)}_{i_1 j_1} O^{(2)}_{i_2 j_2} \dots O^{(d)}_{i_d j_d} \, T_{j_1 j_2 \dots j_d }\, .
\eea
The observables in this model are the contractions of  an even number, say $2n$ with $n \in \mathbb{N}$, of tensors $T$ which are obviously invariant under $\otimes_{a=1}^dO(N_a)$ transformations. 
We simply name them $O(N)$ invariants. 
Such invariants generalize real matrix traces and will be
denoted likewise: 
\be
O_K(T) = \Tr (T ^{2n}) = \sum_{j^{(k)}_l} T_{j^{(1)}_1 j^{(1)}_2 \dots j^{(1)}_d } T_{j^{(2)}_1 j^{(2)}_2 \dots j^{(2)}_d }
\dots  T_{j^{(2n)}_1 j^{(2n)}_2 \dots j^{(2n)}_d }
K(\{j^{(1)}_{l}\} ; \{j^{(2)}_l\}  ; \dots;\{ j^{(2n)}_l\}  )\, ,
\ee
where the kernel $K(\cdot)$ factors in Kronecker delta's  and
identifies the indices of the tensors in a particular pattern; the sole contractions permitted involve
 the tensor indices with identical color labels $i=1\dots d$. 
An elegant way of encoding the contraction pattern of tensors consists in a  $d$-regular graph with edge coloring with $d$ different colors, and one of each color at every vertex (representing each tensor). Calling $b$ the colored graph, the invariant denotes equivalently $O_K(T) = O_b(T) $. We will detail  this in the next section.  

We  build a physical model by introducing a partition function
\bea
Z=\int d\nu(T)\exp(-S_N(T)) \, ,
\eea
where the action $S_N(T)=\sum_{b}\lambda_bN^{-\rho(b)}O_b(T)$ is defined as a finite sum over 
some $O(N)$ tensor invariants
representing the model interactions each with coupling $\lambda_b$ and $\rho(b)$ scaling parameter;
$d\nu(T)$ is a tensor field measure. 

In this work, we will consider only correlators that are Gaussian. This means 
that the field measure will be Gaussian and of the form
\beq
d\nu(T) = \prod_{j_l} dT_{j_1 j_2 \dots j_d } \, e^{-  O_2(T) }\,, 
\qquad
O_2(T) = \sum_{j_k} (T_{j_1 j_2 \dots j_d })^2 \, .
\label{gauss}
\eeq
In other terms, $O_2(T)$ plays the role of a quadratic mass term. 
The free propagator of the Gaussian measure is given by 
\beq\label{propag}
\la  T_{i_1 i_2 \dots i_d} T_{j_1 j_2 \dots j_d} \ra  = \int  d\nu(T) \,  T_{i_1 i_2 \dots i_d} T_{j_1 j_2  \dots j_d}  = \delta_{i_1j_1}\delta_{i_2j_2} \dots \delta_{i_d j_d}, 
\eeq
and will be used in the Wick theorem for computing Gaussian correlators. 
We will be interested in the mean values of observables that are defined 
by 
\bea
\label{correlinit}
&&
\la O_b (T) \ra = \frac{ 1}{\int d\nu (T) } \int d\nu (T) O_b(T) \, , \cr\cr
&&
\la O_b (T)O_{b'} (T)  \ra = \frac{ 1}{\int d\nu (T) } \int d\nu (T) O_b(T)O_{b'} (T)  \, .
\eea
The second correlator will be restricted to normal order
allowing only Wick contractions from $O_{b} (T) $ to $O_{b'} (T) $. 
In section \ref{sect:correl}, enlightened by the symmetric group formulation 
of the $O(N)$ invariants, we will reformulate \eqref{correlinit} and 
analyse the representation algebraic structure brought by the 2pt-correlator. 
The first correlator is sketched as it evaluates by modifying the previous calculation method.

\section{Counting  $O(N)$ invariants}
\label{section:counting}

Counting the number of invariants based on  the contractions of $2n$ copies of 
tensors $T_{i_1, \cdots, i_d}$, starts by a symmetric group  construction. 
Actually, this enumeration problem expresses as a permutation-TFT that 
we also discuss. 
Finally, switching to  representation theory, we derive the same counting 
formula in terms of the famous Kronecker coefficients.

\subsection{Enumeration of  rank $d\geq3$ tensor invariants} 
\begin{figure}[h]
\begin{center} 
     \begin{minipage}[t]{.8\textwidth}
     \centering
\includegraphics[angle=0, width=12cm, height=4.5cm]{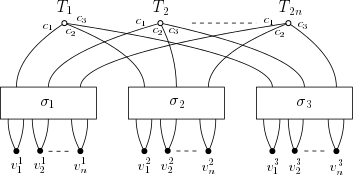}
\end{minipage}
\end{center}
\caption{  Rank $d=3$ orthogonal tensor contractions. } 
\label{fig:coloredgraphs}
\end{figure}

Orthogonal invariants are in one-to-one in correspondence
with  $d$-regular colored graphs (see for instance \cite{CarrooAdri}). Contrary to the graphs corresponding to unitary invariants \cite{Gur4,BenRamg}, the present graphs are not bipartite and, so, might be non-orientable.
It is always possible to make a graph bipartite by inserting another type of vertex
of valence 2 called ``black'' (henceforth the initial  vertices are called ``white'') on each edge of the graph. 
We therefore perform that transformation and the new vertices are denoted $v_i^j$, $i=1, \cdots, 2n$ (recall that $2n$ is the number of tensors) and $j=1, \cdots, d$. The resulting graph is neither regular,  nor properly edge-colored. 
It is however bipartite as illustrated in Figure \ref{fig:coloredgraphs}.  This 
property  concedes a description of a colored graph in  symmetric groups language. 

We shall focus on $d=3$.   The general case $d$ will follow
from this case.  We denote $S_{2n}$ the symmetric group of order $(2n)!$. 
Counting possible graphs consists of enumerating the triples
\beq
( \s_1 , \s_2 , \s_3 ) \in  S_{2n} \times S_{2n}  \times S_{2n}  
\eeq
subjected to the equivalence 
\bea
\label{gamma-equivs} 
( \s_1 , \s_2 , \s_3 ) \sim ( \gamma_1 \s_1 \gamma , \gamma_2 \s_2 \gamma , \gamma_3 \s_3 \gamma ) \,, 
\eea
where  $\gamma \in S_{2n} $ and the $ \gamma_i$ belong to the so-called wreath product  subgroup $\in S_{n}[S_2] \subset S_{2n}$. 
We intend to count the points in the double coset 
\bea\label{doublecoset}
 (  S_{n}[S_2] \times  S_{n}[S_2] \times  S_{n}[S_2] ) \ses  ( S_{2n}  \times S_{2n}  \times S_{2n}  ) / \diag ( S_{2n}  ) \,. 
\eea
Let us denote $ Z_3(2n)$ the cardinality of this double coset. 

In a broader setting \cite{Read}, for two subgroups $ H_1 \subset G $ and $H_2 \subset G $, the cardinality  of the double coset 
$|H_1\ses G / H_2| $ is given by 
\bea \label{hgh2} 
|H_1\ses G / H_2| = 
{ 1 \over |H_1|  | H_2 |   } \sum_{  C } Z_C^{ H_1 \rightarrow G } Z^{ H_2 \rightarrow G }_C \,\Sym ( C ) \,. 
\eea
The sum is over conjugacy classes of $G$, and $Z_C^{ H \rightarrow G }$ is the number of elements of $H$ in the conjugacy class $C$  of $G$.

The conjugacy classes of $S_{2n}  \times S_{2n}  \times S_{2n}  $ are  determined by  triples $(p_1 , p_2 , p_3 )$, where each $p_i$ is a partition   of $2n$. The presence of the subgroup $\diag ( S_{2n}  ) $ implies
that only conjugacy classes determined by a triple 
$( p , p , p )$ should be conserved in the above
sum. 

Applying \eqref{hgh2}, we get 
\bea\label{sol3} 
Z_3(2n) &=& { 1 \over {[n! (2!)^{n}]^3 (2n)!}} \sum_{ p \,\vdash 2n }  
Z_{(p,p,p)}^{(S_{n}[S_2])^{\times 3} \to (S_{2n} )^{\times 3}}
\times \Big( { (2n)! \over \Sym ( p  ) }  \Big)  ( \Sym ( p  ))^3  
\cr\cr
&=&  { 1 \over {[n! (2!)^{n}]^3 }} \sum_{ p \,\vdash 2n }  
Z_{(p,p,p)}^{ (S_{n}[S_2])^{\times 3} \to (S_{2n} )^{\times 3} }
\times   ( \Sym ( p  ))^2 
\cr\cr
&\text{with}& 
\Sym(p):= \prod_{i=1}^n (i^{p_i})(p_i!) \,,
\eea
and 
where the sum over $p= (p_\ell)_\ell $ is performed over all partitions of 
$2n=\sum_i i p_i$.  The cardinality of a conjugacy class $T_p$ of 
$S_{2n} $ with cycle structure determined by a partition $p$ is given 
by $|T_p|= (2n)!/\Sym(p)$.
Next, we must determine the size of $Z_{(p,p,p)}^{(S_{n}[S_2])^{\times 3} \to (S_{2n} )^{\times 3}}$ which is 
\be
Z_{(p,p,p)}^{ (S_{n}[S_2])^{\times 3} \to (S_{2n} )^{\times 3} }= 
(Z_{p}^{ S_{n}[S_2] \to S_{2n}  })^3 \,. 
\ee 
We can get a single factor in this product as 
\be
  { 1 \over {n! (2!)^{n} }} Z_{p}^{ S_{n}[S_2] \to S_{2n} }
   =  \hbox{Coefficient }  [  \cZ_{2}^{S_\infty[S_2]}(t, \vec x) , t^{n} x_1^{p_1} x_2^{p_2}\dots x_{2n} ^{p_{2n} }  ]
   \,, 
\ee
where appears the generating function of  the number of wreath product elements in a certain conjugacy class
$p \vdash 2n$, namely   
\be
\cZ_{d}^{S_\infty[S_d]}(t, \vec x) = \sum_{n} t^n Z^{ S_n[S_d]}(\vec x)
= e^{  \sum_{i=1}^{\infty} \, \frac{t^i}{i}\,  
\Big[\sum_{q \vdash d} \prod_{\ell=1}^{d} 
\big(\frac{ x_{i\ell} }{ \ell } \big)^{\nu_\ell} 
\frac{1}{ \nu_\ell! } \Big] } \,,
\ee
where $\vec x = (x_1, x_2, \dots )$, and  $q= (\nu_\ell)_\ell$ is a partition of $d$, 
such that $\sum_{\ell} \ell \nu_\ell = d$.

The expression \eqref{sol3} finally computes to 
\bea
Z_3(2n) = \sum_{ p \,\vdash 2n }  
\big(\hbox{Coefficient }  [  \cZ_{2}^{S_\infty[S_2]}(t, \vec x) , t^{n} x_1^{p_1} x_2^{p_2}\dots x_{2n} ^{p_{2n} }  ]\big)^3( \Sym ( p  ))^{2} \, . 
\eea
In general, for arbirtrary $d$, the above calculation is straightforward and yields, for any $d\ge 2$, 
\bea\label{gened}
Z_d(2n) = \sum_{ p \,\vdash 2n }  
 \big(\hbox{Coefficient }  [  \cZ_{2}^{S_\infty[S_2]}(t, \vec x) , t^{n} x_1^{p_1} x_2^{p_2}\dots x_{2n} ^{p_{2n} }  ]\big)^d ( \Sym ( p  ))^{d-1} \,. 
\eea
We can generate the sequences $Z_3(2n)$  and  $Z_4(2n)$ (both with $n=1, \cdots, 10$) using a Mathematica program in Appendix \ref{app:mathsage} and obtain, respectively,
\bea\label{eq:z3}
1, 5, 16, 86, 448, 3580, 34981, 448628, 6854130, 121173330
\eea
and
\bea\label{eq:z4}
1, 14, 132, 4154, 234004, 24791668, 3844630928, 809199787472,\cr 220685007519070, 75649235368772418 \, .
\eea
Following Read \cite{Read}, the number $Z_d(2n)$ of $d$-regular colored graphs made with $2n$ vertices is the coefficient of $t^n$ in $\prod_m\Phi_m(t)$,  with $m$ sufficiently large to collect
all  such coefficients,  and where 
\bea
\Phi_m(t)=\left\{
\begin{array}{rl}
\sum_{j=0}^{\infty}\frac{(A_{m/2}(j))^d}{j!m^j}t^{mj/2} & \mbox{ if } k=1, \\\\ \sum_{j=0}^{\infty}\frac{(2j)^{d-1}}{j!^d}\big(\frac{m^{d-2}}{2^k}\big)^jt^{mj} & \mbox{ if } m\mbox{ is odd},
\end{array}\right.
\eea
and the function $A_k(j)$ is related to the $j$-th Hermite polynomial by
$A_k(j)=(i\sqrt{k})^jH_j(\frac{1}{2i\sqrt{k}})$.

We generate the corresponding sequences $Z_3(2n)$, $n=1, \cdots, 10$, and  $Z_4(2n)$, $n=1, \cdots, 10$, using 
a Mathematica program (in Appendix \ref{app:mathsage}) and the results match with \eqref{eq:z3} and \eqref{eq:z4}, respectively.
Hence, both methods yield the same results. 
The sequence \eqref{eq:z3} naturally corresponds to the OEIS sequence A002830 (number of 3-regular edge colored graphs with $2n$ nodes) \cite{oeis}. 
The sequence \eqref{eq:z4} is not yet reported on the OEIS. Hence, the formula \eqref{gened}  generates arbitrary new sequences for each $d>3$.  

\begin{figure}[h]
\begin{center} 
     \begin{minipage}[t]{.8\textwidth}
     \centering
\includegraphics[angle=0, width=14cm, height=12cm]{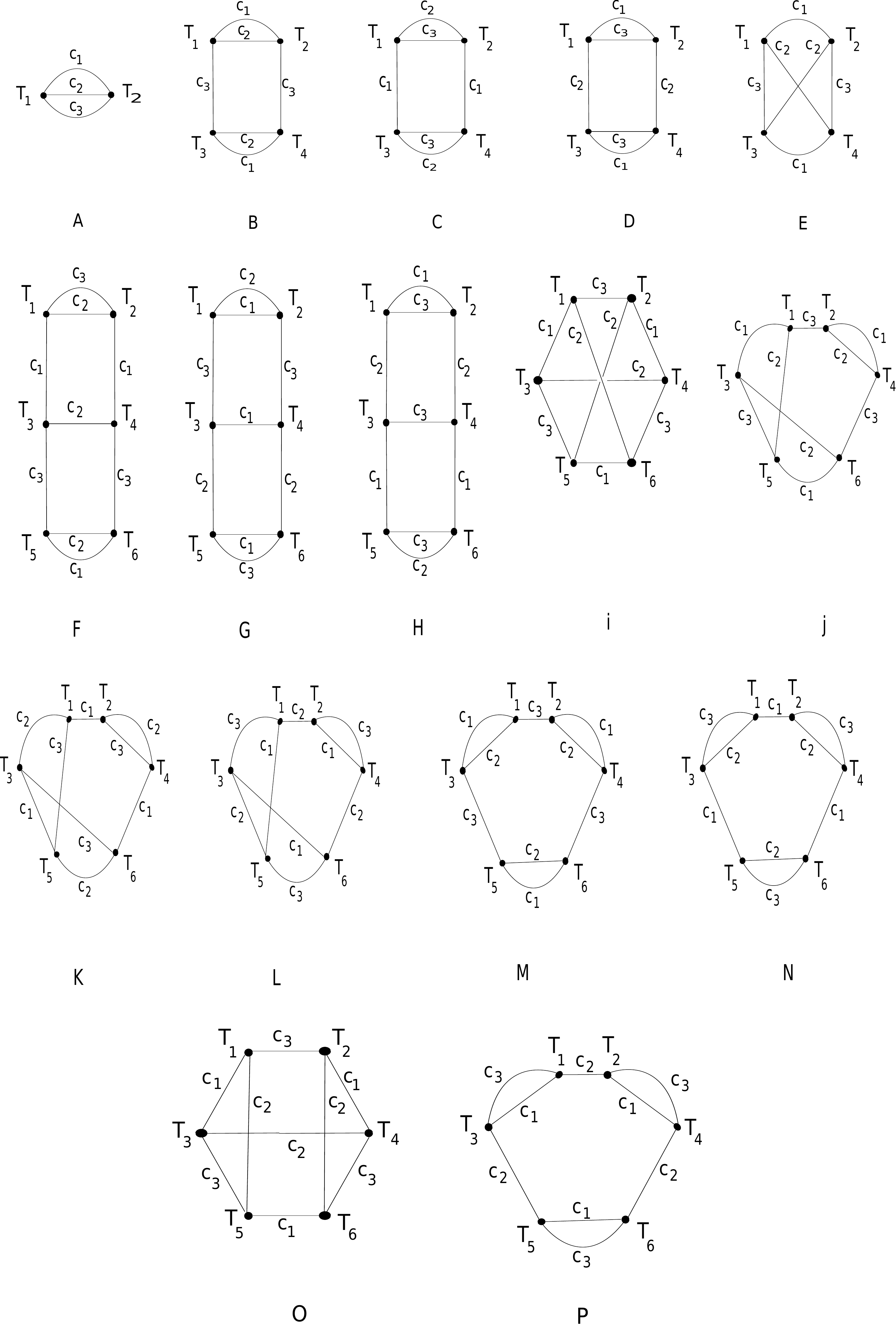}
\end{minipage}
\end{center}
\caption{ Connected colored graphs associated with rank 3 orthogonal tensor invariants
 with up to 6 vertices. } 
\label{fig:coloredgraphs2}
\end{figure}

We must underline that the above counting of observables concerns connected and disconnected graphs (generalized multi-matrix invariants).
To obtain only connnected invariants, we use the plethystic logarithm (Plog) transform on the generating
series of the disconnected invariants. Such a generating function 
also easily programs with the M\"obius $\mu$-function. We obtain the enumeration of connected invariants (see Appendix \ref{app:mathsage}) for rank $d=3$ and $4$, respectively, up to order $n=10$ as,  
\be
1, 4, 11, 60, 318, 2806, 29359, 396196, 6231794, 112137138\,, 
\ee
and 
\bea
1, 13, 118, 3931, 228316, 24499085, 3816396556, 805001547991, 219822379032704, 
\cr
75417509926065404 \,.
\eea
As an illustration,  Figure \ref{fig:coloredgraphs2} depicts the rank 3 connected orthogonal invariants up to valence 6. The colors $c_i=1,2,3$,  should be permuted to generate the full set of 
connected invariants. 


\subsection{TFT formulation}
From the above symmetric group formulation of the counting of tensor invariants, one  extracts more information
via other correspondences.  
In particular, the enumeration reformulates as a partition function of a Topological Field Theory on
a 2-complex (in short TFT$_2$)  with $S_{2n}$ and its subgroup $S_n[S_2]$ as gauge groups. 
For a review of TFT's, see \cite{CMR1,CMR2} and, in notation closer to what we aim at, see \cite{FeynCount, Gal}. 

\begin{figure}[h]
 \centering
     \begin{minipage}[t]{.9\textwidth}
      \centering
\includegraphics[angle=0, width=11cm, height=3.5cm]{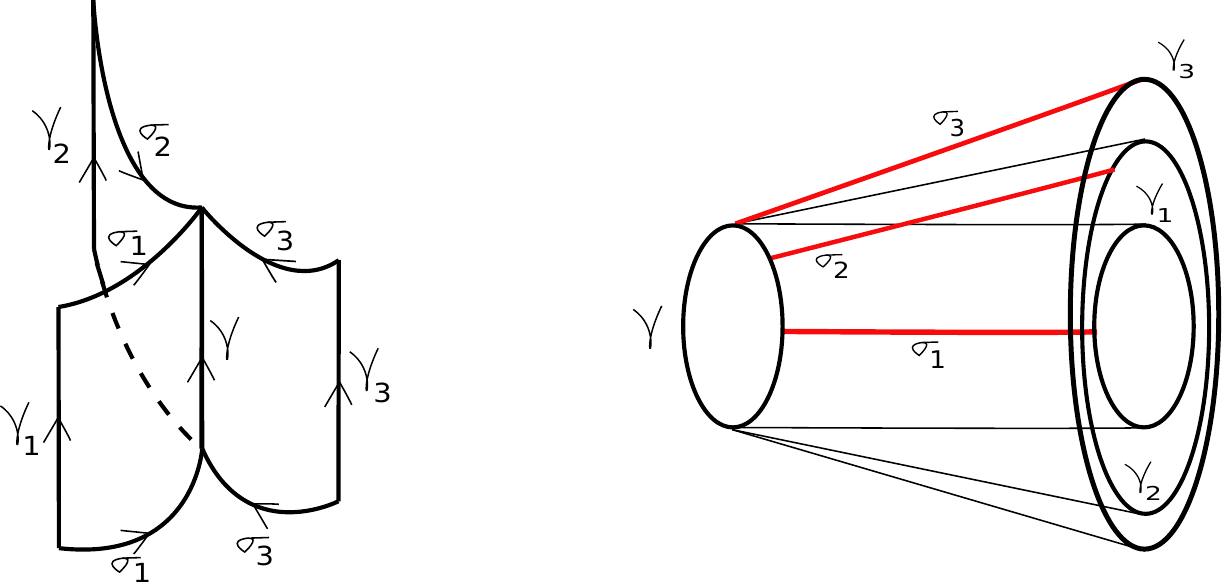}
\caption{ {\small TFT$_2 $ associated with the counting of orthogonal invariants. }}
\label{fig:tft2Orth}
\end{minipage}
\end{figure}

Consider the counting of classes in the double coset \eqref{doublecoset},
denote it as $ Z_3(2n)$, and then consider the relation \eqref{hgh2}.
Using Burnside's lemma, we have in standard notations: 
\be\label{BL}
Z_3(2n)  =  { 1 \over {[n! (2!)^{n}]^3 (2n)!}}  
\sum_{\gamma_i \in S_n[S_2]}\sum_{\s_i \in S_{2n}}
\sum_{\gamma\in S_{2n}} \delta(\gamma_1 \s_1 \gamma \s_1^{-1}) \delta(\gamma_2 \s_2 \gamma \s_2^{-1})
\delta(\gamma_3 \s_3 \gamma  \s_3^{-1})\,, 
\ee
where $\delta$ is the Kronecker symbol  on $S_{2n}$. 
This counting interprets as a partition function of a TFT$_2$ on a cellular complex given 
by Figure \ref{fig:tft2Orth}. On that lattice, we use two gauge groups $S_{2n}$ and  $S_n[S_2]$. The topology of that
2-complex is that of three cylinders sharing the same end circle. Thus, enumerating orthogonal invariant
corresponds to a $S_{2n}$--TFT$_2$ on  3 glued cylinders along one circle, with a restriction
of the gauge group to be $S_n[S_2]$ at the opposite boundary circle. This TFT$_2$ has boundary holonomies
endowed with $S_n[S_2]$ group elements. 

\begin{figure}[h]
 \centering
     \begin{minipage}[t]{.9\textwidth}
      \centering
\includegraphics[angle=0, width=12cm, height=3.5cm]{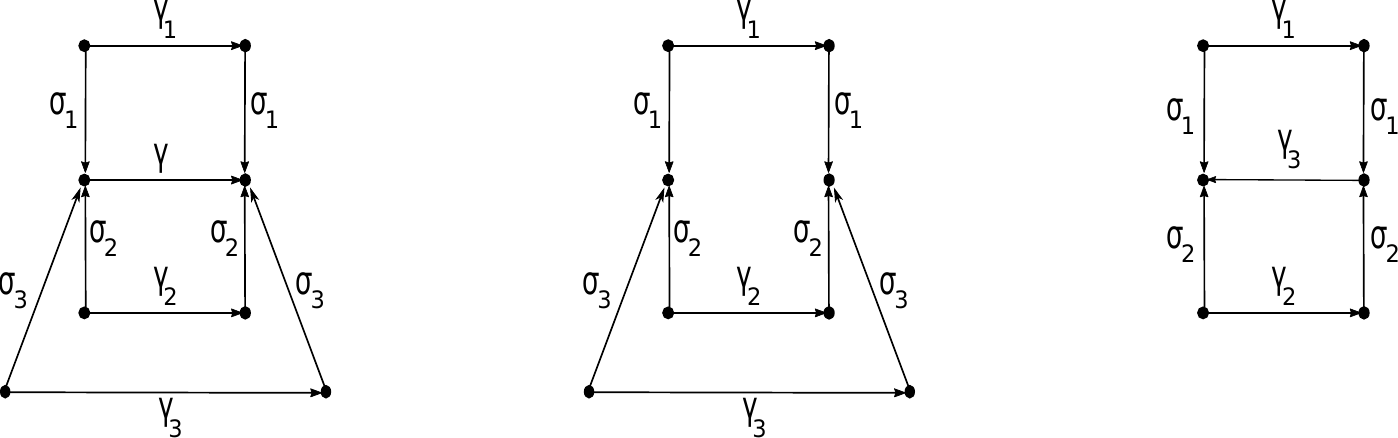}
\caption{ {\small Topological transformations of the  2-complex leaving the partition function stable.}}
\label{fig:toptrf}
\end{minipage}
\end{figure}

By successively integrating some delta functions, the TFT$_2$ formulation produces alternative interpretations of the same counting. We extract $\gamma$ from  \eqref{BL}
and get $\gamma =\s_3^{-1} \gamma^{-1}_3 \s_3   $
such that 
\bea
Z_3(2n)  =  { 1 \over {[n! (2!)^{n}]^3 (2n)!}}  
\sum_{\gamma_i \in S_n[S_2]}\sum_{\s_i \in S_{2n}}
&&\delta(\gamma_1 \s_1 (\s_3^{-1} \gamma^{-1}_3 \s_3   ) \s_1^{-1})\times \cr 
&&\delta(\gamma_2 \s_2 (\s_3^{-1} \gamma^{-1}_3 \s_3   ) \s_2^{-1}) \, . 
\eea
A change of variables $\s_{1,2} \to \s_{1,2} \s_3^{-1} $ leads us to
\be\label{eq:gamma}
Z_3(2n) 
= { 1 \over {[n! (2!)^{n}]^3}}  
\sum_{\gamma_i \in S_n[S_2]}\sum_{\s_{1,2} \in S_{2n}}
\delta(\gamma_1 \s_1  \gamma_3 \s_1^{-1}) \delta(\gamma_2 \s_2  \gamma_3 \s_2^{-1}) \, . 
\ee
This integration illustrates, in Figure \ref{fig:toptrf}, as the removal of a 1-cell associated
with the variable $\gamma$ in the 2-complex. The partition function therefore shows two types
of invariances: the extraction of $\gamma$ corresponds to one type of topological invariance, and then,
 it is followed by the change of variables $\s_{1,2} \to \s_{1,2} \s_3^{-1} $ corresponding to a  topological invariance 
 of a second kind. 

\begin{figure}[h]
 \centering
     \begin{minipage}[t]{.9\textwidth}
      \centering
\includegraphics[angle=0, width=8cm, height=2.5cm]{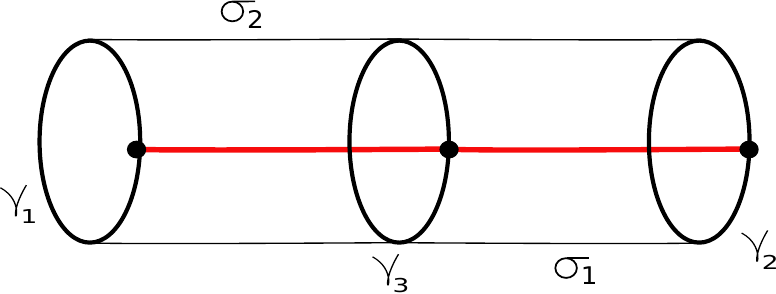}
\caption{ {\small Cylinder with 3 defects.}}
\label{fig:defects}
\end{minipage}
\end{figure}

Thus,  the partition function \eqref{eq:gamma} can also be written as 
\be
\label{eq:heffect}
Z_3(2n)  =  Z(S^1\times I; (D_{ S_n[S_2] })^{\times 3}) \, ,
\ee
where $Z(S^1\times I; (D_{ S_n[S_2] })^{\times 3})$ is the partition function obtained by inserting 3 $S_n[S_2]$-defects, one at each end of the cylinder $S^1\times I$, and another one at finite
time $t_0 \in I$, see Figure \ref{fig:defects}. A defect defines as a closed non-intersecting loop with a marked point. 
The relation \eqref{eq:heffect} shows that orthogonal  invariants 
are in one-to-one correspondence with $n$-fold covers of the cylinder with 3 defects, 
up to  a (symmetry) factor, the stabilizer subgroup of the graph
that we denote ${\rm Aut}(G_{\s_1, \s_2,\s_3})$. 

The order of the stabilizer infers from 
\bea
\Sym(\s_1, \s_2) =  \sum_{\gamma_i \in S_n[S_2]}
\delta(\gamma_1 \s_1  \gamma_3 \s_1^{-1}) \delta(\gamma_2 \s_2  \gamma_3 \s_2^{-1})
= {\rm Aut}(G_{\s_1, \s_2,\s_3})
\eea
which  also relates to the number of equivalences
 $(S_n[S_2] \times S_n[S_2] )\ses (S_n \times S_n) / \diag(S_n[S_2]) $ 
 corresponding to a fixed $(\s_1, \s_2)$. 

The TFT formulation of the counting could enrich it with a geometrical picture. 
Most of the time, the base space of the TFT is viewed as a string worldsheet. 
The counting becomes now counting of worldsheet maps over a cylinder with defects. 
As noticed elsewhere \cite{BenRamg,BenSamj}, 
this once again shows that a link may exist between tensor models and string theory, 
which could be elucidated via the TFT formalism. 
Such link may be worth investigating in the future.

\

\noindent{\bf Rank $d$ counting and TFT$_2$ --} 
More generally, for rank $d\ge 3$, the counting $Z_d(2n)$ has a TFT$_2$ formulation that
generalizes what we discuss  above in a straightforward manner: 
\bea\label{BLd}
Z_d(2n)  &=&  { 1 \over {[n! (2!)^{n}]^d (2n)!}}  
\sum_{\gamma_i \in S_n[S_2]}\sum_{\s_i \in S_{2n}}
\sum_{\gamma\in S_{2n}} \prod_{i=1}^d \delta(\gamma_i \s_i \gamma \s_i^{-1})\crcr
&=& { 1 \over {[n! (2!)^{n}]^d}}  
\sum_{\gamma_i \in S_n[S_2]}\sum_{\s_{i} \in S_{2n}}
\prod_{i=1}^{d-1}
\delta(\gamma_i \s_i  \gamma_{d} \s_i^{-1}) \, . 
\eea
The first equation of \eqref{BLd} shows that, in rank $d$, the TFT$_2$-formulation of the counting
extends Figure \ref{fig:tft2Orth} as the gluing of $d$ cylinders along one circle. 
After integration, the second equations reveals that the counting orthogonal invariants amounts therefore counting of weighted covers of $d-1$ cylinders with 
$d$ defects, with one of the defects shared by all cylinders. In formula, we have
$Z_d(2n)  =  Z(S^1\times I; (D_{ S_n[S_2] })^{\times d})$.

\subsection{The counting as a Kronecker sum}
\label{rev}

We now revisit the counting \eqref{BL} under a different light, that of the  representation theory of the
symmetric group (Appendix \ref{app:SRT} reviews the main identities used in this section
and the following). Irreducible representations (irreps) of the symmetric group $S_{2n}$ are labeled by 
partitions $R \vdash 2n$, that are also Young diagrams. 

Starting from the Burnside lemma  formulation of the counting \eqref{BL}, 
consider the following expansion of the counting of rank 3 invariants using the representation 
theory of $S_{2n}$: 
\bea
\label{eq:dimK} 
&& Z_3(2n)  = \frac{1}{[n!(2!)^n]^3(2n)!}\sum_{\gamma_l \in S_n[S_2]} \sum_{\sigma_l \in S_{2n}} \sum_{\gamma \in S_{2n}} \delta(\gamma_1 \sigma_1 \gamma \sigma_1^{-1})
\delta(\gamma_2 \sigma_2 \gamma \sigma_2^{-1})\delta(\gamma_3 \sigma_3 \gamma \sigma_3^{-1}) \cr\cr
&& = \frac{1}{[n!(2!)^n]^3(2n)!} \sum_{\gamma_l \in S_n[S_2]} \sum_{\gamma \in S_{2n}} \sum_{R_l \,\vdash\, 2n} \chi^{R_1}(\gamma_1) \chi^{R_1}(\gamma) \chi^{R_2}(\gamma_2) \chi^{R_2}(\gamma) \chi^{R_3}(\gamma_3) \chi^{R_3}(\gamma) \crcr
&& = \frac{1}{[n!(2!)^n]^3}  \sum_{R_l \,\vdash\, 2n} \textbf{C}(R_1,R_2,R_3) \Big[ \sum_{\gamma_1 \in S_n[S_2]} \chi^{R_1}(\gamma_1)\Big]\Big[ \sum_{\gamma_2 \in S_n[S_2]} \chi^{R_2}(\gamma_2)\Big]
\eea
where  $ \chi^{R}(\cdot)$ denotes the character in the representation $R$; 
we used the identity \eqref{delgsg0} in Appendix \ref{app:reptheor} to compute the
delta's, and the Kronecker coefficient is defined as
\bea
\textbf{C}(R_1,R_2,R_3) = \frac{1}{2n!} \sum_{\gamma \in S_{2n}}\chi^{R_1}(\gamma) \chi^{R_2}(\gamma)
\chi^{R_3}(\gamma) \,. 
\eea
The Kronecker defines the multiplicity of the representation $R_3$ in the tensor product
$R_1 \otimes R_2$, or the multiplicity of the trivial representation in $R_1 \otimes R_2 \otimes R_3$
when expanded back in irreps. 

Above, the sums over the subgroup $S_n[S_2]$ have been not yet performed. 
To proceed with these sums, we will use a useful result by Howe \cite{Howe} (see also
a result by Mizukawa, proposition 4.1  in \cite{mizukawa}, 
and also \cite{macdo,Ivanov} or a more recent use of it in \cite{Caputa:2013hr}): 
\be
\sum_{\gamma \in S_n[S_2]} \chi^{R}(\gamma)
 = |S_n[S_2]|  \, m^{R}
\ee
where $m^{R}= 1$ if   $R$ is an ``even'' partition, that is, all 
its  row lengths are even, and $m^{R}=0$ otherwise. 
This result is derived from a more general formula 
$\sum_{\gamma \in S_n[S_2]} \chi^{A}(\gamma)\chi^{R}(\gamma)
 = |S_n[S_2]|  \, m^{A|R}$, where  $A$ is an irreps of $S_n[S_2]$ subduced by $R$
 irreps of $S_{2n}$, and then inserting $A$ as the trivial representation $\overline{[2n]}$ of $S_n[S_2]$.  

Then, we obtain, inserting this in \eqref{eq:dimK} 
\be\label{kroneven}
Z_3(2n)  =  \sum_{R_l \,\vdash\, 2n/ R_l \text{ is even }} \textbf{C}(R_1,R_2,R_3) \,. 
\ee
Comparing this sequence and \eqref{eq:z3}, 
we produce a Sage code (see Appendix \ref{app:mathsage}) showing that the numbers generated by \eqref{kroneven} match with   \eqref{eq:z3}.

In the next section, we will show that, this number is also the dimension of an algebra $\cK_{3}(n)$.
It is an interesting problem to investigate how the counting of colored graphs could 
contribute to the famous problem of giving a combinatorial interpretation to the
Kronecker coefficients \cite{iken1,Blasiak} (in the same way that Littlewood-Richardson coefficient
have found a combinatorial description). From previous work \cite{BenSamj}, we know that
the sum of squares of Kronecker coefficients associated with $S_n$ 
equals the number of $d$-regular bipartite colored graphs made with $n$ black and
$n$ white vertices. Here the interpretation is the following, the number of  $d$-regular 
colored graphs (not necessarily bipartite) equals the sum of all Kronecker's 
precluded those that are defined with partitions with odd rows. 
An idea to contribute to the above problem is to refine the counting of
graphs in a way to boil down to a single Kronecker coefficient. 
In other words, given a non vanishing Kronecker coefficient 
is it possible to list all graphs contributing to that Kronecker coefficient? 
This is certainly a difficult problem that will require new  tools 
in representation theory.

\medskip 

\noindent
\textbf{Counting rank-$d$ tensor invariants -} The above counting  generalize quite naturally at any rank $d$ as
\begin{equation}
\begin{aligned}
Z_d(2n) & = \frac{1}{[n!(2!)^n]^d(2n)!} \sum_{\gamma_l \in S_n[S_2]} \sum_{\sigma_l \in S_{2n}} \sum_{\gamma \in S_{2n}} \prod_{i=1}^d \delta(\gamma_i \sigma_i \gamma \sigma_i^{-1}) \\
& = \frac{1}{[n!(2!)^n]^d} 
 \sum_{R_l \,\vdash\, 2n}\sum_{\gamma_l \in S_n[S_2]} \textbf{C}_d(R_1,\ldots, R_d) \chi^{R_1}(\gamma_1) \ldots \chi^{R_d}(\gamma_d) \\
 & = \sum_{R_l \,\vdash\, 2n/ R_l \text{ is even } }\sum_{\gamma_l \in S_n[S_2]} \textbf{C}_d(R_1,\ldots, R_d) 
 \label{countingZ3n}
\end{aligned} 
\end{equation}
where we introduced the  notation
\be
\textbf{C}_k(R_1,\ldots, R_k)= \frac{1}{(2n)!} \sum_{\gamma \in S_{2n}}\chi^{R_1}(\gamma) \ldots \chi^{R_k}(\gamma)\,. 
\ee
This counts the multiplicity of the one dimensional trivial $S_{2n}$ irrep in the tensor product of irreps $R_1 \otimes \ldots \otimes R_k$. It expresses as a convoluted product of Kronecker coefficients as
\be
\textbf{C}_k(R_1,\ldots, R_k) = \sum_{S_l \vdash 2n } \textbf{C}(R_1,R_2,S_1) \left[\prod_{i=1}^{k-4} \textbf{C}(S_i,R_{i+2}, S_{i+1})\right] \textbf{C}(S_{k-3},R_{k-1},R_k)\,. 
\ee


\section{Double coset algebra}
\label{sect:doublcoset}

We now discuss the underlying structure, an algebra,  determined by the counting of the $O(N)$ invariants. 
The rank 3 case is first addressed for the sake of simplicity, and from that, we will infer the general rank-$d$ case whenever possible. 

Consider $\mathbb{C}[S_{2n}]$, the group algebra of $S_{2n}$. Our construction depends on 
tensor products of that space. 

\medskip 
\noindent{\bf $\cK_d(2n)$ \textbf{as a double coset algebra in} $\mathbb{C}[S_{2n}]^{\otimes d}$ -} 
We fix $d=3$. 
Consider  $\sigma_1 \otimes \sigma_2 \otimes \sigma_3$ as an element of the group algebra $\mathbb{C}[S_{2n}]^{\otimes 3}$,  and three left actions of 
the subgroup $S_n[S_2]$ and the diagonal right action $\diag(\mathbb{C}(S_{2n}))$ on this triple as:
\begin{equation}
\sigma_1 \otimes \sigma_2 \otimes \sigma_3 \rightarrow \sum_{\gamma_i \in S_n[S_2]} \sum_{\gamma \in S_{2n}} 
\gamma_1 \sigma_1 \gamma \otimes  \gamma_2 \sigma_2 \gamma \otimes  \gamma_3 \sigma_3 \gamma\,.
\end{equation}
$\cK_3(2n)$ is the vector subspace of $\mathbb{C}[S_{2n}]^{\otimes 3}$ which is invariant under these subgroup actions:
\begin{equation}
\label{Kunalg}
\cK_3(2n)= \text{Span}_{\mathbb{C}} \left\{ \sum_{\gamma_i \in S_n[S_2]} \sum_{\gamma \in S_{2n}} \gamma_1 \sigma_1 \gamma \otimes  \gamma_2 \sigma_2 \gamma \otimes  \gamma_3 \sigma_3 \gamma, \quad \sigma_1, \sigma_2, \sigma_3 \in S_{2n} \right\}
\end{equation}
It is obvious that $\dim \cK_3(2n) = Z_3(2n)$, since each base element represents the graph equivalent class counted once in $Z_3(2n)$. 
Pick two base elements, called henceforth graph base elements, and consider their product
\bea\label{multip}
&&
\Big[\sum_{\gamma_i \in S_n[S_2]} \sum_{\gamma \in S_{2n}} \gamma_1 \sigma_1 \gamma \otimes  \gamma_2 \sigma_2 \gamma \otimes  \gamma_3 \sigma_3 \gamma \Big] \Big[ \sum_{\tau_i \in S_n[S_2]} \sum_{\tau \in S_{2n}} \tau_1 \s'_1 \tau \otimes  \tau_2 \s'_2 \tau \otimes  \tau_3 \sigma_3' \tau \Big] \cr\cr
&& =  \sum_{\gamma_i,\tau_i  \in S_n[S_2]} \sum_{\gamma,\tau \in S_{2n}} 
\gamma_1 \sigma_1 \gamma \tau_1 \s'_1 \tau  \otimes  \gamma_2 \sigma_2 \gamma \tau_2 \s'_2 \tau \otimes  \gamma_3 \sigma_3 \gamma\tau_3 \sigma_3' \tau \cr\cr
&&
=\sum_{\tau_i \in S_n[S_2] }\sum_{\gamma \in S_{2n}}  \Big[\sum_{\gamma_i \in S_n[S_2]}
 \sum_{\tau \in S_{2n}} \gamma_1 (\sigma_1 \gamma \tau_1 \s'_1) \tau  \otimes  \gamma_2 (\sigma_2 \gamma \tau_2 \s'_2) \tau \otimes  \gamma_3 (\sigma_3 \gamma\tau_3 \sigma_3' )\tau \Big] 
\eea
This shows that the multiplication remains in the vector space. 
Hence,  $\cK_3(2n)$ is an algebra and \eqref{multip} defines a graph multiplication. 
The proof is totally similar for $\cK_d(2n)$ (considering $d$ factors in the tensor product) which is thus an algebra of dimension $Z_d(2n)$. 

\medskip 

\begin{figure}[h]
 \centering
     \begin{minipage}[t]{.9\textwidth}
      \centering
\includegraphics[angle=0, width=15cm, height=7cm]{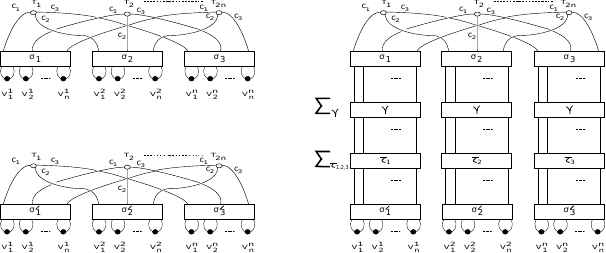}
\caption{ {\small Product of two graph base elements (on the left) gives a sum of graphs (on the right).}}
\label{fig:graphcompo}
\end{minipage}
\end{figure}

The product of graphs in the  algebra $\cK_3(2n)$ illustrates as in Figure \ref{fig:graphcompo}. 

\medskip 

\noindent{\bf Gauge fixing -} There is a gauge fixing procedure in the construction of orthogonal 
invariants. One initially fixes a permutation $\s_i$ but is still able to generate all 
invariants. Consider $\xi = (12)(34)\dots (2n-1,2n)$, and we fix $\s_1$ to belong to 
the stabilizer of $\xi$, i.e.  $\s_1^{-1} \xi  \s_1 = \xi$.  Since the Stab$_\xi= S_n[S_2]$, 
we simply mean that we choose $\s_1$ to be in that subgroup. 
We already observe  a difference with the  unitary case \cite{BenSamj}. 
Indeed, while the gauge fixing in the unitary case leads to the definition of a permutation 
centralizer algebra, the gauge fixing here will not bring such an algebra. 
The main difference with the unitary case also rests on the fact that
the left and right invariances on the triple  $(\sigma_1 , \sigma_2 , \sigma_3) $ in this case are radically different.

\medskip

\noindent{\bf Associativity -}
In the graph base,  we can check the associativity of the product of elements of $\cK_{3}(2n)$: 
\bea 
&& \Bigg(\Big[\kun{\gamma}{\sigma_1}{\sigma_2}{\sigma_3}{\gamma}\Big]
\Big[ \kun{\tau}{\sigma_1'}{\sigma_2'}{\sigma_3'}{\tau} \Big]\Bigg)\crcr
&& \times \Big[ \kun{\alpha}{\sigma_1''}{\sigma_2''}{\sigma_3''}{\alpha} \Big]\crcr
&&
 = \sum_{\tau_i,\, \alpha_i } \sum_{\gamma,\, \tau}
\Big[\sum_{\gamma, \alpha} 
\gamma_1 \sigma_1\gamma\tau_1\sigma_1'\tau\alpha_1\sigma_1''\alpha \otimes 
\gamma_2 \sigma_2\gamma\tau_2\sigma_2'\tau\alpha_2\sigma_2'' \alpha \otimes
 \gamma_3\sigma_3\gamma\tau_3\sigma_3'\tau\alpha_3\sigma_3'' \alpha\Big]\crcr
&& = \Big[\kun{\gamma}{\sigma_1}{\sigma_2}{\sigma_3}{\gamma}\Big] \crcr
&& \times \left( \Big[ \kun{\tau}{\sigma_1'}{\sigma_2'}{\sigma_3'}{\tau} \Big]
\Big[ \kun{\alpha}{\sigma_1''}{\sigma_2''}{\sigma_3''}{\alpha} \Big] \right)\crcr
&&
\eea

The proof easily extends to any $d$, and we therefore claim the following: 
\begin{proposition}
$\cK_{d}(2n)$ is an associative unital sub-algebra of  $\mathbb{C}[S_{2n}]^{\otimes 3}$. 
\end{proposition}
The unity is given by the equivalence class of $(id,id,id)$.  Such element corresponds to the 
disconnected graph made with $n$  connected components with full contraction of $n$ pairs of tensors. 

\ 

\noindent{\bf Pairing -} There is an inner product (that we will call pairing) on $\cK_d(2n)$ defined from the linear 
extension of the delta function from the symmetric group to the tensor product group algebra (see   \eqref{bdel} in Appendix \ref{app:groualg} for details pertaining to the following notation). Take two base elements (in obvious notation) and evaluate using proper change of variables: 
\bea
&&
\bdel\Big( \sum_{\g_i,\g}  \otimes_i^d \g_i \s_i \g ; \sum_{\tau_i,\tau}  \otimes_i^d \tau_i \s'_i \tau \Big) = 
\sum_{\g_i,\g} \sum_{\tau_i,\tau}  \prod_i^d \delta( \g_i \s_i \g(\tau_i \s'_i \tau )^{-1})
\crcr
&&
 = [(2n)!(n!2^n)]\sum_{\g_i,\g} \prod_i^d \delta( \g_i \s_i \g ( \s'_i)^{-1} ) \,. 
\eea
Thus, either the tuples $(\s_1,\s_2, \dots, \s_d)$ and $(\s'_1,\s'_2, \dots, \s'_d)$ define equivalent graphs
$G_{\s_1,\s_2, \dots, \s_d}$ and $ G_{\s'_1,\s'_2, \dots, \s'_d}$, respectively, 
or the result is 0. This precisely tells us that the graph base forms an orthogonal system.  
The above computes further  using the order of the automorphism group of the graph
\be
\bdel\Big( \sum_{\g_i,\g}  \otimes_i^d \g_i \s_i \g ; \sum_{\tau_i,\tau}  \otimes_i^d \tau_i \s'_i \tau \Big) = 
  [(2n)!(n!2^n)]\delta(G_{\s_1,\s_2, \dots, \s_d}; G_{\s'_1,\s'_2, \dots, \s'_d}) {\rm Aut}(G_{\s_1,\s_2, \dots, \s_d}) \,. 
\ee
Therefore, there exists a non degenerate bilinear pairing on $\cK_d(2n)$ and 
the following holds: 
\begin{theorem}\label{theosemisimple}
$\cK_d(2n)$ is an associative  unital semi-simple algebra. 
\end{theorem}

As a corollary of Theorem \ref{theosemisimple}, the Wedderburn-Artin theorem guarantees that $\cK_d(2n)$  decomposes in matrix subalgebras. 
It might be interesting to investigate a base of such a decomposition of $\cK_d(2n)$ in 
irreducible  matrix subalgebras. One could be tempted to think that, at $d=3$, restricting to $\cK_3(2n)$,  the Kronecker coefficients for even partitions could be themselves squares, and therefore define  the dimensions of the irreducible subalgebras. This is not the case
as this can be easily shown using the same Sage code given in Appendix \ref{app:mathsage}
(by printing the Kronecker). This point is postponed for future investigations. 
In the mean time, it is legitimate to ask a representation base with labels  that reflect the dimension \eqref{kroneven}. This is the purpose of the next paragraph.

\

\noindent{\bf Constructing a representation theoretic  base $Q$ of $\cK_3(2n)$ -}
Let us introduce the representation base of $\mathbb{C}[S_{2n}]$ given by the elements 
\be \label{Foub}
Q^{R}_{ij}  = \frac{\kappa_R}{(2n)!}\sum_{\s\in S_{2n}} D^{R}_{ij}(\s) \s  \,,\qquad \text{with}\; \kappa_R^2 = (2n)! d(R)\,, 
\ee
that obey the orthogonality relation  $\delta(Q^{R}_{ij} ; Q^{R'}_{i'j'} )= \delta_{RR'}\delta_{ii'}\delta_{jj'}$. 
The base $\{Q^{R}_{ij}  \}$ counts $\sum_{R \vdash 2n} (d(R))^2= (2n)!$ elements and forms the Fourier theoretic  base of $\IC[S_{2n}]$. Appendix \ref{app:groualg} collects 
a few other properties of this base for a general permutation group.

We fix $d=3$ and build now the invariant representation theoretic (Fourier for short) base of the algebra  $\cK_3(2n)$ \eqref{Kunalg}.  
Consider the right diagonal action $\rho_R(\cdot )$  and the three left actions $\varrho_i(\cdot )$ on the tensor product $\mathbb{C}[S_{2n}]^{\otimes 3}$.
Then we write:
\begin{equation}\begin{aligned}
& \sum_{\substack{\gamma_1,\,\gamma_2,\,\gamma_3 \in S_n[S_2]}} \sum_{\gamma \in S_{2n}} \varrho_1(\gamma_1)\varrho_2(\gamma_2)\varrho_3(\gamma_3) \rho_R(\gamma) \, Q^{R_1}_{i_1 j_1}\otimes Q^{R_2}_{i_2 j_2}\otimes Q^{R_3}_{i_3 j_3} \\
& = \sum_{\gamma_a} \sum_{\gamma} \gamma_1\, Q^{R_1}_{i_1 j_1}\gamma \otimes \gamma_2\, Q^{R_2}_{i_2 j_2}\, \gamma \otimes\gamma_3 \,Q^{R_3}_{i_3 j_3} \,\gamma \\
& = \sum_{\gamma_a} \sum_{\gamma} \sum_{p_l \,, q_l}
D^{R_1}_{p_1 i_1}(\gamma_1) Q^{R_1}_{p_1 q_1}D^{R_1}_{j_1 q_1}(\gamma) \otimes
D^{R_2}_{p_2 i_2}(\gamma_2) Q^{R_2}_{p_2 q_2}D^{R_2}_{j_2 q_2}(\gamma) \otimes
D^{R_3}_{p_3 i_3}(\gamma_3) Q^{R_3}_{p_3 q_3}D^{R_3}_{j_3 q_3}(\gamma) \\
& = \frac{(2n)!}{d(R_3)} \sum_{\gamma_a} \sum_{p_l \,, q_l}\sum_{\tau} C^{R_1,R_2;R_3,\tau}_{j_1,j_2;j_3}
C^{R_1,R_2;R_3,\tau}_{q_1,q_2;q_3} D^{R_1}_{p_1 i_1}(\gamma_1) 
D^{R_2}_{p_2 i_2}(\gamma_2) D^{R_3}_{p_3 i_3}(\gamma_3)
Q^{R_1}_{p_1 q_1} \otimes Q^{R_2}_{p_2 q_2} \otimes Q^{R_3}_{p_3 q_3}. 
\end{aligned}\end{equation}
We used \eqref{tauq} to multiply group elements with the $Q$ base, see Appendix  \ref{app:groualg};
then use \eqref {lem2:DDD=CC} to sum over $\g$ the 3 representation matrices, see in Appendix \ref{app:cgc}.

We couple this last result with a Clebsch-Gordan coefficient, in order to get, using \eqref{dddc=c}:
\begin{equation}\begin{aligned}
& \sum_{j_l} C^{R_1,R_2;R_3,\tau}_{j_1,j_2;j_3}\sum_{\gamma_a} \sum_{\gamma} \varrho_1(\gamma_1)\varrho_2(\gamma_2)\varrho_3(\gamma_3) \rho_R(\gamma) \, Q^{R_1}_{i_1 j_1}\otimes Q^{R_2}_{i_2 j_2}\otimes Q^{R_3}_{i_3 j_3} 
\label{sumddd}\\
& = (2n)! \sum_{p_l \,, q_l} C^{R_1,R_2;R_3,\tau}_{q_1,q_2;q_3} 
\sum_{\gamma_1} D^{R_1}_{p_1 i_1}(\gamma_1)
\sum_{\gamma_2} D^{R_2}_{p_2 i_2}(\gamma_2)
\sum_{\gamma_3} D^{R_3}_{p_3 i_3}(\gamma_3)
Q^{R_1}_{p_1 q_1} \otimes Q^{R_2}_{p_2 q_2} \otimes Q^{R_3}_{p_3 q_3} \,.
\end{aligned}\end{equation}
Once again, we should stress  the fact that $\sum_{\gamma \in S_n[S_2]} D^{R}_{p q}(\gamma) \neq 0$,
if and only if $R$ is a partition of $2n$ with even rows. This condition will be always assumed in 
the next calculations. 
Now, we can split the Wigner matrix element using branching coefficients of  
$S_n[S_2]$ in $S_{2n}$. Consider $V^R$ a irreps  $S_{2n}$ (see Appendix \ref{app:SRT} listing a few  basic facts on representation theory of $S_n$ and our notations), and the subgroup inclusion $S_n[S_2] \subset S_{2n}$, we can decompose $V^R$ in irreps  $V^r$  of $S_n[S_2]$ as
\bea
V^R = \oplus_{r} V^{r}\otimes V_{R,r}
\eea
where $ V_{R,r}$ is a vector space of dimension the multiplicity of the irreps $r$ in $R$. 
A state in this decomposition denotes $\ket{ r, m_r, \nu_r }$, 
where $m_r$ labels the states of $V^{r}$ and $\nu_r = 1, \dots, \dim V_{R,r}$. 

The branching coefficients that are of interest 
are the coefficients of $\ket{ r, m_r, \nu_r }$ when decomposed in an 
orthonormal base of the irreps $R$: 
\bea
B^{R;\, r, \nu_r}_{i; \,m_r}  = 
\langle R, i \ket{ r, m_r, \nu_r }  = 
\langle r , m_r, \nu_r\ket{  R, i  }  \, . 
\eea
The last relation is deduced from the fact that we use real 
representations. Using the decomposition of the identity, 
the branching coefficients satisfy the following identities
\bea
&&
\sum_{ i}
B^{R;\, r, \nu_r}_{i; \,m_r} 
 B^{R;\, s, \nu_s}_{i; \,m_s} = \delta_{rs}\delta_{\nu_r \nu_s}  \delta_{m_r m_s} 
 \label{orthobranch0} \\
 &&
\sum_{r,m_r, \nu_r}
B^{R;\, r, \nu_r}_{i; \,m_r} 
 B^{R';\, r, \nu_r}_{i'; \,m_r} = \delta_{RR'}\delta_{ii'} \,. 
 \label{orthobranch}
\eea
We have the following useful relation, for $\s \in S_n[S_2]$, 
\bea
\sum_{j} D^R_{ij}(\s) \,B^{R; \,r, \nu_r}_{j; m_r}  = 
\sum_{m'_r} D^r_{m_rm_r'}(\s) \,B^{R; \,r, \nu_r}_{i; m'_r} \, ,
\eea
where $ D^r_{m_rm_r'}(\s)$ is the representation matrix of $\s$ as an element of $S_n[S_2]$. 
Restricting this to $r= \overline{[2n]}$, the trivial representation of $S_n[S_2]$ that
is one dimensional and with multiplicity always 1 for all $R$, we obtain: 
\bea
\sum_{j} D^R_{ij}(\s) \,B^{R; \overline{[2n]}, 1}_{j; 1}  = 
D^{\overline{[2n]}}_{11}(\s) \,B^{R; \overline{[2n]}, 1}_{i;1}  =  B^{R; \overline{[2n]}, 1}_{i;1}\, . 
\label{brancto1}
\eea

We now treat  the sum over the representation matrices in \eqref{sumddd}. Inserting twice a complete set of states therein, we get 
\bea
\sum_{\s\in S_{n}[S_2]} D^{R}_{ij}(\s) = 
\sum_{\s\in S_{n}[S_2]}
\sum_{r,\nu_r,m_r; s,\nu_s,m_s}  B^{ R; \, r, \nu_r}_{i;\, m_r} 
B^{ R; \, s, \nu_s}_{j;\, m_s} \langle r,\nu_r,m_r | \s | 
  s,\nu_s,m_s \rangle   \,. 
\eea 
Noting that $\sum_{s\in S_{n}[S_2]} \s = \sum_{s\in S_{n}[S_2]} \s \chi^{\overline{[2n]}}(\s)$
is, up to the factor  $1/[n! 2^n]$,  nothing but the projector onto the trivial $\overline{[2n]}$ representation of $S_n[S_2]$, 
the overlap computes to 
\bea
\sum_{\s\in S_{n}[S_2]} \langle r,\nu_r,m_r | \s | 
  s,\nu_s,m_s \rangle = (2^n n!) \delta_{r,\overline{[2n]}}
  \delta_{s,\overline{[2n]}} \delta_{1 m_r} \delta_{1m_s}\delta_{1\nu_s} \delta_{1\nu_r} \,, 
  \eea
 since we have
 \bea
 &&
 \sum_{\s\in S_{n}[S_2]} \s | 
  s,\nu_s,m_s \rangle   = 
   \sum_{\s\in S_{n}[S_2]}  \chi^{\overline{[2n]}}(\s) \sum_{k} D^s_{m_sk }(\s) | 
  s,\nu_s,k  \rangle \cr\cr
  &&
   = \sum_{\s\in S_{n}[S_2]}  D^{\overline{[2n]}}_{11}(\s) \sum_{k} D^s_{m_sk }(\s) | 
  s,\nu_s,k  \rangle
   = \frac{2^n n!}{d(\overline{[2n]})}
 \sum_{k} \delta_{\overline{[2n]} ,s} \delta_{1m_s} \delta_{1\nu_s}\delta_{1k}   | 
  s,\nu_s,k  \rangle\cr\cr
  &&
  =  (2^n n!)
  \delta_{\overline{[2n]} ,s} \delta_{1m_s}\delta_{1\nu_s}  | 
  \overline{[2n]},1,1  \rangle \,. 
 \eea
 Hence, 
\be\label{DSS}
\sum_{\s\in S_{n}[S_2]} D^{R}_{ij}(\s) = 
(2^n n!)
B^{ R; \,  tr}_{i} B^{R;\,  tr}_{j}  \,, 
\ee
where we have defined $B^{ R; \,  tr}_{i} =  \langle R,i \,|
\overline{[2n]},1,1 \rangle $. 

From the above calculation, we finally get from \eqref{sumddd}: 
\begin{equation}
\begin{aligned}
& 
\sum_{j_l} C^{R_1,R_2;R_3,\tau}_{j_1,j_2;j_3}\sum_{\gamma_a} \sum_{\gamma} \varrho_1(\gamma_1)\varrho_2(\gamma_2)\varrho_3(\gamma_3) \rho_R(\gamma) \, Q^{R_1}_{i_1 j_1}\otimes Q^{R_2}_{i_2 j_2}\otimes Q^{R_3}_{i_3 j_3} \\
& = (2n)! ( n!2^n) ^3 B^{R_1;tr}_{i_1}B^{R_2;tr}_{i_2}B^{R_3;tr}_{i_3} \sum_{p_l \,, q_l} C^{R_1,R_2;R_3,\tau}_{q_1,q_2;q_3} 
B^{R_1; \, tr}_{p_1} 
B^{R_2; \,  tr}_{p_2} 
B^{R_3;\,  tr}_{p_3} 
Q^{R_1}_{p_1 q_1} \otimes Q^{R_2}_{p_2 q_2} \otimes Q^{R_3}_{p_3 q_3} 
\end{aligned}
\end{equation}
We now define an element  
\bea \label{unbasis}
&&
Q^{R_1,R_2,R_3,\tau} 
 = \kappa_{\vec{R}}  
 \sum_{p_l \, ,q_l} C^{R_1,R_2;R_3,\tau}_{q_1,q_2;q_3} 
B^{R_1; \,  tr}_{p_1} 
B^{R_2;\,  tr}_{p_2} 
B^{R_3; \,  tr}_{p_3} 
Q^{R_1}_{p_1 q_1} \otimes Q^{R_2}_{p_2 q_2} \otimes Q^{R_3}_{p_3 q_3} \cr\cr
&& 
=  \kappa_{\vec{R}}  
\frac{\kappa_{R_1}\kappa_{R_2}\kappa_{R_3}}{((2n)!)^3} 
\sum_{\s_i} \sum_{p_l \, ,q_l} C^{R_1,R_2;R_3,\tau}_{q_1,q_2;q_3} 
\Big[\prod_{i=1}^3 B^{R_i; \,  tr}_{p_i} 
D^{R_i}_{p_i q_i}(\s_i)  \Big]
\s_1 \otimes \s_2 \otimes \s_3
\label{basis}
\eea
where $\kappa_{\vec{R}}$ is a normalization constant to be fixed later and
the notation $\vec R$  stands for $(R_1, R_2, R_3)$. The set $\{Q^{R_1,R_2,R_3,\tau}\}$ is of cardinality 
the counting of  orthogonal invariants given by \eqref{kroneven}.

\ 
 
\noindent{\bf Invariance -}
Let us check that the element $Q^{R_1,R_2,R_3,\tau}$ is invariant under 
left multiplication on each factor and diagonal right multiplication: 
\bea 
&&
(\gamma_1 \otimes \gamma_2 \otimes \gamma_3   )
Q^{R_1,R_2,R_3,\tau}  ( \gamma   \otimes \gamma  \otimes \gamma ) 
 = 
\kappa_{\vec{R}}   
 \sum_{p_l \, ,q_l} C^{R_1,R_2;R_3,\tau}_{q_1,q_2;q_3} 
B^{R_1; \,  tr}_{p_1} 
B^{R_2;\,  tr}_{p_2} 
B^{R_3; \,  tr}_{p_3}  \cr\cr
 && \times
\sum_{\ell_1, j_1} D^{R_1}_{ \ell_1 p_1}(\gamma_1)  Q^{R_1}_{\ell_1 j_1}  D^{R_1}_{ q_1 j_1}(\gamma) \otimes
\sum_{\ell_2, j_2} D^{R_2}_{ \ell_2 p_2}(\gamma_2)  Q^{R_2}_{\ell_2 j_2} D^{R_2}_{q_2 j_2}(\gamma) \otimes 
\sum_{\ell_3, j_3} D^{R_3}_{ \ell_3 p_3}(\gamma_3) Q^{R_3}_{\ell_3 j_3} D^{R_3}_{q_3 j_3}(\gamma)\crcr
&& 
 = 
\kappa_{\vec{R}}   
 \sum_{j_l}  C^{R_1,R_2;R_3,\tau}_{j_1,j_2;j_3}  
   \sum_{p_l, \ell_l} 
D^{R_1}_{ \ell_1 p_1}(\gamma_1)  B^{R_1; \,  tr}_{p_1} 
D^{R_2}_{ \ell_2p_2}(\gamma_2)  B^{R_2;\,  tr}_{p_2} 
D^{R_3}_{ \ell_3 p_3}(\gamma_3)  B^{R_3; \,  tr}_{p_3}  \; 
 Q^{R_1}_{\ell_1 j_1}   \otimes
  Q^{R_2}_{\ell_2 j_2} \otimes 
 Q^{R_3}_{\ell_3 j_3}  \crcr
&& 
 = \kappa_{\vec{R}}   
 \sum_{j_l, \ell_l }  C^{R_1,R_2;R_3,\tau}_{j_1,j_2;j_3}  
  B^{R_1; \,  tr}_{\ell_1} 
B^{R_2;\,  tr}_{\ell_2} 
  B^{R_3; \,  tr}_{\ell_3}  \; 
 Q^{R_1}_{\ell_1 j_1}   \otimes
  Q^{R_2}_{\ell_2 j_2} \otimes 
 Q^{R_3}_{\ell_3 j_3} 
  = Q^{R_1,R_2,R_3, \tau}\,, 
\eea
where we used once again  \eqref{tauq} and \eqref{dddc=c} at  intermediate step and  the identity  \eqref{brancto1} to get the last line.

We check a few properties of the product of elements of
$\cK_{3}(2n)$. 

\

\noindent{\bf Product -} The 
elements \eqref{Foub} of the Fourier base of $\mathbb{C}[S_{2n}]$ multiply as follows
(see Appendix \ref{app:groualg}.) 
\bea \label{qqprod}
Q^R_{ij} Q^{R'}_{kl} = \frac{\kappa_R}{d(R)}\delta_{RR'}\delta_{jk} Q^{R'}_{il}\,.
\eea 
The definition \eqref{unbasis} and  relation \eqref{qqprod}  allow us to compute the product 
\beq \begin{aligned}
& \quad Q^{R_1,R_2,R_3,\tau} Q^{R_1',R_2',R_3',\tau'} \\
&= \frac{\kappa_{\vec{R}}\kappa_{\vec{R'}}\kappa_{R_1}\kappa_{R_2}\kappa_{R_3}}{d(R_1)d(R_2)d(R_3)}\delta_{\vec{R}\vec{R'}} \sum_{p_l \, q_l \, a_l \, b_l} 
C^{R_1,R_2;R_3,\tau}_{q_1,q_2;q_3} C^{R_1',R_2';R_3',\tau'}_{b_1,b_2;b_3} \\
& \qquad \times 
B^{R_1; \,  tr}_{p_1} 
B^{R_2;\,  tr}_{p_2} 
B^{R_3; \,  tr}_{p_3}
B^{R_1'; \,  tr}_{a_1} 
B^{R_2';\,  tr}_{a_2} 
B^{R_3'; \,  tr}_{a_3}
Q^{R_1'}_{p_1 b_1} \otimes
Q^{R_2'}_{p_2 b_2} \otimes 
Q^{R_3'}_{p_3 b_3} \delta_{q_1a_1} \delta_{q_2a_2} \delta_{q_3a_3}\\
& = \frac{\kappa_{\vec{R}}\kappa_{R_1}\kappa_{R_2}\kappa_{R_3}}{d(R_1)d(R_2)d(R_3)}\delta_{\vec{R}\vec{R'}} \Big[\sum_{q_l} C^{R_1',R_2';R_3',\tau}_{q_1,q_2;q_3} 
B^{R_1'; \,  tr}_{q_1} 
B^{R_2';\,  tr}_{q_2} 
B^{R_3'; \,  tr}_{q_3} \Big] Q^{R_1',R_2',R_3',\tau'}. 
\end{aligned} 
\eeq
Hence, the product of  two base elements expands in terms of $Q^{R_1,R_2,R_3,\tau}$.  
In a compact notation, we write
\beq
Q^{R_1,R_2,R_3,\tau}
Q^{R_1',R_2',R_3',\tau'} = \delta_{\vec{R}\vec{R'}}  k(\vec{R'},\tau) Q^{R_1',R_2',R_3',\tau'}. 
\eeq
which shows that the product is almost orthogonal. Still it cannot represent the base of Wedderburn-Artin matrix decomposition. The base $\{Q^{R_1,R_2,R_3,\tau}\}$ therefore decomposes $\cK_3(2n)$ in  blocks mutually 
orthogonals in the labels $R_1,R_2,R_3$. Still in each block the decomposition remains unachieved.

\
 
\noindent{\bf Associativity -}  
We check the associativity of the product in the $Q$-base. On the one hand, we have 
\beq \begin{aligned}
&\quad \left(\Qun{} \Qun{'}\right) \Qun{''} \\
& = \frac{\kappa_{\vec{R}}\kappa_{R_1}\kappa_{R_2}\kappa_{R_3}}{d(R_1)d(R_2)d(R_3)}\delta_{\vec{R}\vec{R'}} \Big[\cbbb{'}{}\Big] \\
& \quad \times \frac{\kappa_{\vec{R'}}\kappa_{R_1'}\kappa_{R_2'}\kappa_{R_3'}}{d(R_1')d(R_2')d(R_3')}\delta_{\vec{R'}\vec{R''}} \Big[\cbbb{''}{'}\Big] \Qun{''}
\end{aligned} \eeq
and on the other,
\beq \begin{aligned}
& \quad \Qun{} \left(\Qun{'} \Qun{''}\right) \\
& = \frac{\kappa_{\vec{R'}}\kappa_{R_1'}\kappa_{R_2'}\kappa_{R_3'}}{d(R_1')d(R_2')d(R_3')}\delta_{\vec{R'}\vec{R''}} \Big[\cbbb{''}{'}\Big] \\
& \quad \times \frac{\kappa_{\vec{R}}\kappa_{R_1}\kappa_{R_2}\kappa_{R_3}}{d(R_1)d(R_2)d(R_3)}\delta_{\vec{R}\vec{R''}} \Big[\cbbb{''}{}\Big] \Qun{''}.
\end{aligned} \eeq 
The two expressions are identical. 

\

\noindent{\bf Pairing -} We use the pairing on  $\mathbb{C}[S_{2n}]^{\otimes 3}$
along  the lines \eqref{pairing} and evaluate: 
\beq \begin{aligned}
& \quad \boldsymbol{\delta}(Q^{R_1,R_2,R_3,\tau}; Q^{R_1',R_2',R_3',\tau'})\\
& = \kappa_{\vec{R}}\kappa_{\vec{R'}} \sum_{p_l \, q_l \, a_l \, b_l} 
C^{R_1,R_2;R_3,\tau}_{q_1,q_2;q_3} C^{R_1',R_2';R_3',\tau'}_{b_1,b_2;b_3} 
B^{R_1; \,  tr}_{p_1} 
B^{R_2;\,  tr}_{p_2} 
B^{R_3; \,  tr}_{p_3}
B^{R_1'; \,  tr}_{a_1} 
B^{R_2';\,  tr}_{a_2} 
B^{R_3'; \,  tr}_{a_3}\\
& \hspace*{30mm} \times 
\boldsymbol{\delta}(Q^{R_1}_{p_1 q_1}  \otimes
Q^{R_2}_{p_2 q_2}  \otimes 
Q^{R_3}_{p_3 q_3} ;
Q^{R_1'}_{a_1 b_1}\otimes  Q^{R_2'}_{a_2 b_2} \otimes Q^{R_3'}_{a_3 b_3})\\
& = \kappa_{\vec{R}}\kappa_{\vec{R'}} \sum_{p_l \, q_l \, a_l \, b_l} 
C^{R_1,R_2;R_3,\tau}_{q_1,q_2;q_3} C^{R_1',R_2';R_3',\tau'}_{b_1,b_2;b_3} 
B^{R_1; \,  tr}_{p_1} 
B^{R_2;\,  tr}_{p_2} 
B^{R_3; \,  tr}_{p_3}
B^{R_1'; \,  tr}_{a_1} 
B^{R_2';\,  tr}_{a_2} 
B^{R_3'; \,  tr}_{a_3}\\
& \hspace*{30mm} \times \delta_{\vec{R}\vec{R'}} \delta_{p_1a_1} \delta_{p_2a_2} \delta_{p_3a_3} \delta_{q_1b_1} \delta_{q_2b_2} \delta_{q_3b_3}\\
& = \kappa_{\vec{R}}^2\, d(R_3) 
\sum_{p_l} \Big[ \prod_{i=1}^3 B^{R_i; \,  tr}_{p_i} \Big]^2
\delta_{\vec{R}\vec{R'}} \delta_{\tau \tau'} =  \kappa_{\vec{R}}^2\, d(R_3) \delta_{\vec{R}\vec{R'}} \delta_{\tau \tau'} \, 
\end{aligned}
 \eeq 
 where, in the first line, we used \eqref{pairing},  in the last, \eqref{cc=del}, 
 and the fact that, by \eqref{orthobranch0}, the following holds  $\sum_{p}  [B^{R; \,  tr}_{p} ]^2  =\sum_{p}  \langle
\overline{[2n]},1,1 | \, R,p \rangle \langle R,p \,|
\overline{[2n]},1,1 \rangle   = 1$, for all $R\vdash 2n$. 
We could therefore fix the normalization 
$ \kappa_{\vec{R}}^2\,   = 1/ d(R_3) $.  

The following statement holds: 
\begin{proposition}
$\{Q^{R_1,R_2,R_3,\tau}\}$ is an invariant orthonormal base of $\cK_{3}(2n)$. 
\end{proposition}
\noindent{\bf Proof.}
It is sufficient to show that the graph base expands in terms of the $Q$-base. 
We hold the non degenerate pairing $\boldsymbol{\delta}$ and express 
 any graph base element $G_{\s_1,\s_2, \s_3}$  $= \sum_{\gamma_i \in S_n[S_2]} \sum_{\gamma \in S_{2n}} \gamma_1 \sigma_1 \gamma \otimes  \gamma_2 \sigma_2 \gamma \otimes  \gamma_3 \sigma_3 \gamma$  as
\bea
G_{\s_1,\s_2, \s_3} = \sum_{R_l, \tau} \boldsymbol{\delta}(Q^{R_1,R_2,R_3,\tau}; G_{\s_1,\s_2, \s_3})  Q^{R_1,R_2,R_3,\tau}
\eea 
The definition of $Q^{R_1,R_2,R_3,\tau}$ calls a linear combination of triples $(\tau_1 \otimes   \tau_2 \otimes  \tau_3)$
that must have a non trivial overlap with  $G_{\s_1,\s_2, \s_3}$. Let us compute the overlap between 
the bases. Start with \eqref{basis} and then write (using \eqref{dddc=c} and then \eqref{brancto1})
\bea
\boldsymbol{\delta}(Q^{R_1,R_2,R_3,\tau}; G_{\s_1,\s_2, \s_3}) = 
 \kappa_{\vec{R}}  
\frac{\kappa_{R_1}\kappa_{R_2}\kappa_{R_3}}{((2n)!)^3} 
((2^n n!))^3 (2n!)
 \sum_{a_l \, ,b_l}  C^{R_1,R_2;R_3,\tau}_{b_1,b_2;b_3} 
\Big[\prod_{i=1}^3 B^{R_i; \,  tr}_{a_i} 
D^{R_i}_{a_i b_i}( \sigma_i  )  \Big] . \crcr
\label{QG}
\eea
This number  is, up to the normalization $((2^n n!))^3 (2n!)$, the coefficient of the triple $(\s_1 \otimes \s_2 \otimes \s_3)$ in $Q^{R_1,R_2,R_3,\tau}$. 

\qed 

\medskip

We note that the base $\{Q^{R_1,R_2,R_3,\tau} \}$ is of the correct cardinality, that of $Z_3(2n)$ as
we sought. 

Finding of the Wedderburn-Artin matrix base of $\cK_3(2n)$ means  that
$Z_3(2n)$ can be written as a sum of squares. Interestingly, 
within the TFT$_2$ formulation of the counting, we note that 
the partition function \eqref{eq:gamma} computes further using \eqref{delgsg0} 
as  
\bea\label{eq:gamma2}
Z_3(2n)  &=&  { 1 \over {[n! (2!)^{n}]^3}}  
\sum_{R_{l} \vdash2n}
\Big( \sum_{\gamma_1}\chi^{R_1}(\gamma_1) \Big)
\Big( \sum_{\gamma_2} \chi^{R_2}(\gamma_2 )  \Big)
\Big( \sum_{\gamma_3}  \chi^{R_1}(\gamma_3)  \chi^{R_2}(\gamma_3 )  \Big) \crcr
&=& { 1 \over { n! (2!)^{n} }}  
\sum_{R_{l} \vdash2n \;  / \; R_l \, \text{even} } 
\sum_{\gamma_3}  \chi^{R_1}(\gamma_3)  \chi^{R_2}(\gamma_3 ) \crcr
&=&{ 1 \over { n! (2!)^{n} }}   \sum_{\gamma_3} 
\Big(  \sum_{R \vdash2n \;  / \; R \, \text{even} } 
  \chi^{R}(\gamma_3) \Big)^2 \, , 
  \eea
  thus,  as a normalized sum of squares. This shows that $Z_3(2n)$ could admit
  several decompositions in squares. 
If  $\Big(  \sum_{R \vdash2n \;  / \; R \, \text{even} }   \chi^{R}(\gamma_3) \Big)^2$ 
  is the dimension of a  subalgebra  (given that the characters are integers via the Munurghan-Nakayama rule), 
  this would mean that this decomposition in sub-algebras would be
  labeled by $\gamma_3$ and will be even different from the Wedderburn-Artin decomposition. 
This decomposition deserves further clarification in the present $O(N)$ setting. 

\

\noindent{\bf About projectors --}
Let us define the normalized projectors as 
\begin{equation}
P_1^{S_n[S_2]} P_2^{S_n[S_2]} P_3^{S_n[S_2]} P_R^{S_{2n}} = \frac{1}{[n!(2!)^n]^3 (2n)!} \sum_{\gamma_l \in S_n[S_2]} \sum_{\gamma \in S_{2n}} \varrho_1(\gamma_1)\varrho_2(\gamma_2)\varrho_3(\gamma_3) \rho_R(\gamma) \, , 
\end{equation}
and  check that the trace of their product yields the dimension of the algebra $\cK_3(2n)$: 
\begin{equation}
\dim \cK_3(2n) = \operatorname{tr}_{\mathbb{C}[S_{2n}]^{\otimes 3}}(P_1^{S_n[S_2]} P_2^{S_n[S_2]} P_3^{S_n[S_2]} P_R^{S_{2n}}) = \tr_{\cK_3(2n)}(\mathds{1}) \, . 
\end{equation}
We have
\begin{align}
&\quad  \sum_{\gamma_a \in S_n[S_2]} \sum_{\gamma \in S_{2n}} \varrho_1(\gamma_1)\varrho_2(\gamma_2)\varrho_3(\gamma_3) \rho_R(\gamma) \, Q^{R_1}_{i_1 j_1}\otimes Q^{R_2}_{i_2 j_2}\otimes Q^{R_3}_{i_3 j_3}\nn \\
& = \sum_{\gamma_a} \sum_{\gamma} \sum_{p_l \, q_l}
D^{R_1}_{p_1 i_1}(\gamma_1) D^{R_1}_{j_1 q_1}(\gamma)
D^{R_2}_{p_2 i_2}(\gamma_2) D^{R_2}_{j_2 q_2}(\gamma) 
D^{R_3}_{p_3 i_3}(\gamma_3)D^{R_3}_{j_3 q_3}(\gamma) \nn\\
& \quad \times Q^{R_1}_{p_1 q_1} \otimes Q^{R_2}_{p_2 q_2} \otimes  Q^{R_3}_{p_3 q_3} \,. 
\end{align}
To compute the trace, pair this with $Q^{R_1}_{i_1 j_1}\otimes Q^{R_2}_{i_2 j_2}\otimes Q^{R_3}_{i_3 j_3}$ using the orthonormality property $\delta(Q^R_{ij};Q^S_{kl}) = \delta_{RS} \delta_{ik} \delta_{jl}$ and sum over $R_l, \, i_l,\, j_l$ yielding
\begin{align}
& \quad \sum_{R_l \vdash S_{2n}} \sum_{\gamma_a} \sum_{\gamma} \sum_{p_l \, q_l, i_l, j_l}
D^{R_1}_{p_1 i_1}(\gamma_1) D^{R_1}_{j_1 q_1}(\gamma)
D^{R_2}_{p_2 i_2}(\gamma_2) D^{R_2}_{j_2 q_2}(\gamma) 
D^{R_3}_{p_3 i_3}(\gamma_3)D^{R_3}_{j_3 q_3}(\gamma) \nn \\
& \quad \times \delta_{i_1 p_1}  \delta_{j_1q_1}\delta_{i_2p_2}\delta_{j_2q_2} \delta_{i_3p_3}\delta_{j_3q_3} \nn \\
& = \sum_{R_l \vdash S_{2n}} \sum_{\gamma_a} \sum_{\gamma} \sum_{ i_l, j_l}
D^{R_1}_{i_1 i_1}(\gamma_1) D^{R_1}_{j_1 j_1}(\gamma)
D^{R_2}_{i_2 i_2}(\gamma_2) D^{R_2}_{j_2 j_2}(\gamma) 
D^{R_3}_{i_3 i_3}(\gamma_3)D^{R_3}_{j_3 j_3}(\gamma) \nn \\
& = (2n)! \sum_{R_l \vdash S_{2n}} \sum_{\gamma_a} \textbf{C}(R_1,R_2,R_3) \chi^{R_1}(\gamma_1) \chi^{R_2}(\gamma_2) \chi^{R_3}(\gamma_3) \, . 
\end{align}
Hence we find \eqref{eq:dimK}  using Burnside's lemma, and we have $Z_3(n)=\dim \cK_3(2n)$.

\section{Correlators}
\label{sect:correl}

Let us analyze Gaussian correlators, starting with $d=3$ and then extending it at any $d$. 
We consider the normal ordered correlator of two observables $O_b (T)O_{b'} (T)$ in the Gaussian measure $d\nu (T)$  \eqref{gauss}. Normal order means that we only allow contraction from $O_b (T)$ to $O_{b'} (T)$.

\

\noindent{\bf Rank $d=3$ correlator -}
Before computing the correlators, a few remarks must be done.
 A 3-tuple of permutations labels the observables:  $O_b (T) = O_{\s_1, \s_2, \s_3} (T)$ and $O_{b'} (T)=  O_{\tau_1, \tau_2, \tau_3} (T)$.  Recall that an observable $O_{\s_1, \s_2, \s_3}(T) $ is in fact defined by a contraction of
 tensor indices. This contraction pattern, that gives in return
the color edges of the graph associated with the observable, is not defined by 
the triple $(\s_1, \s_2, \s_3)$ but by the following triple 
\bea
(\ws_1, \ws_2, \ws_3) = 
(\s_1^{-1} \xi \s_1, \s_2^{-1} \xi  \s_2, \s_3^{-1} \xi \s_3),
\eea 
where we recall that  $\xi $ is the fixed permutation $(12)(34)\dots (2n-1,2n)$. The justification of this is immediate:
each swop in $\xi$ corresponds to a label of the half-lines of the vertex $v^i_j$, see Figure \ref{fig:coloredgraphs}. 
Consider the $l$-th edge of color $i$ from the $l$-th tensor. 
The vertex links $v^i_j$ the image of $\s_i(l)$ and the pre-image through $\s_i$ of $\xi(\s_i(l))$. 
We need the following convenient notation for tensors:  $T_{a_{i1}a_{i2}a_{i3}}$, 
the index $i=1,\dots,2n$ stands for the label of the tensor which at the end will not matter
in the definition of the observable. Using this, an observable made of 
the contraction of $2n$ tensors can be expressed as: 
\bea\label{observ}
O_{\s_1, \s_2, \s_3} (T) = \sum_{a_{ij}} \prod_{i=1}^{2n} \prod_{j=1}^{3} 
\delta^{a_{ij}}_{a_{\ws_j(i)j}} \prod_{i=1}^{2n}T_{a_{i1}a_{i2}a_{i3}}
\eea
where $a_{ij}= 1, \dots, N$.  There are many redundant  Kroneckers $\delta$ in the previous
 expression. However,  the calculus here is discrete and so there are no particular issues. 
When we will compute the correlator using the Wick theorem, it is the triple 
$(\ws_1, \ws_2, \ws_3) $ that is concerned.

The Wick contraction between two observables, in the normal order, introduces a permutation $\mu\in S_{2n}$. 
A  correlator simply counts cycles of a convolution of permutations. Let us determine which convolution is
that, using twice \eqref{observ} and the free propagator \eqref{propag}: 
 \bea
\label{correl2}
&&
\la O_{\s_1, \s_2, \s_3} (T)O_{\tau_1, \tau_2, \tau_3} (T)  \ra = \frac{ 1}{\int d\nu (T) } \int d\nu (T) O_{\s_1, \s_2, \s_3}(T)O_{\tau_1, \tau_2, \tau_3} (T)  \crcr
&&
 = 
 \sum_{\mu} \sum_{a_{ij}, b_{kl}} 
  \Big[\prod_{i=1}^{2n} \prod_{j=1}^{3} 
\delta^{a_{ij}}_{a_{\ws_j(i)j}} \delta^{b_{ij}}_{b_{\wt_j(i)j}}\Big] 
  \Big[\prod_{i=1}^{2n}\prod_{j=1}^3 \delta^{a_{ij}}_{b_{\mu(i)j}}\Big] \,. 
\eea
Summing over the $b_l$ variables and using a change of variable, $b_{ij}= a_{\mu^{-1}(i)j}$,  lead us to 
\bea
&&
\la O_{\s_1, \s_2, \s_3} (T)O_{\tau_1, \tau_2, \tau_3} (T)  \ra = 
\sum_{\mu} \sum_{a_{ij}} 
  \Big[\prod_{i=1}^{2n} \prod_{j=1}^{3} 
\delta^{a_{ij}}_{a_{\ws_j(i)j}}
\delta^{a_{\mu^{-1}(i)j} }_{  a_{\mu^{-1}\wt_j(i)j}   }\Big]  \crcr
&&
 = 
\sum_{\mu} \sum_{a_{ij}} 
  \Big[\prod_{i=1}^{2n} \prod_{j=1}^{3} 
\delta^{a_{ij}}_{a_{\ws_j(i)j}}
\delta^{a_{ij} }_{  a_{\mu^{-1}\wt_j \mu (i)j}   }\Big] 
=
\sum_{\mu} \sum_{a_{ij}} 
  \Big[\prod_{i=1}^{2n} \prod_{j=1}^{3} 
\delta^{a_{ij}}_{a_{\ws_j(i)j}}
\delta^{a_{ij} }_{  a_{\mu^{-1}\wt_j \mu \ws(i)j}   }\Big] \,,
\eea
where we also used $\ws_j ^{-1}= \ws_j $. 
We already guess that the correlator expresses as a power of $N$  in a number of cycles
of $\mu^{-1}\wt_j \mu \ws_j$. However, the proof is not obvious 
because of the redundancy of the $\delta$ introduced in the definition of the observable, 
see \eqref{observ}. 

The following statement holds 
\begin{lemma}\label{lem1}
Let $a_i $ be an integer, $a_i=1,\dots, N$, for $i=1,\dots, 2n$. Then, 
(at fixed color $j$ that we will omit in the ensuing notation) 
\bea\sum_{a_{i}} 
\Big[
\prod_{i=1}^{2n}
\delta^{a_{i}}_{a_{\ws(i)}}
\delta^{a_{i} }_{  a_{\mu^{-1}\wt \mu \ws(i)}   }\Big]
 = N^{\cy( \mu^{-1}\wt \mu \ws )}  \, , 
\label{cycles} 
\eea
where  $\cy(\s)$ is the number of cycles of the permutation $\s$. 
\end{lemma}

\proof 
The sole issue here is the redundancy of the Kronecker's. In fact, 
there is enough information in the above sum to withdraw the correct number of cycles. 
Call ``vertex $\delta$'s'' those appearing in the product $\prod_{i=1}^{2n} \delta^{a_{i}}_{a_{\ws(i)}}$, 
and (Wick) ``contraction  $\delta$'s'' the remaining ones coming from the resolution of the Wick contraction. 
Note there are  redundancies in each product of $\delta$'s. 

Consider a fixed index $i$: to make things easy, we start by the simple case given by  $\mu^{-1}\wt \mu \ws (i)=i$. 
 If $\mu^{-1}\wt \mu \ws ^{-1}(i) = i$,
then $(i)$ is a 1-cycle of $\mu^{-1}\wt \mu \ws $ and we also have
$\ws(i)= \mu^{-1}\wt \mu(i)$. Thus, we have, among the contraction $\delta$'s , 2 distinct $\delta$'s which become trivial 
$\delta^{a_i}_{a_i}$ and  $\delta^{a_{\ws(i)}}_{a_{\ws(i)}}$. 
The sums over $a_i$ and  $a_{\ws(i)}$ boil down to a single 
sum precisely because of  the vertex $\delta^{ a_{i}} _{ a_{\ws(i)}} $. Hence that cycle is counted once. 

Let us inspect the general case. For an arbitrary $i$, 
call  $q_i \ge 1$ the smallest integer such that $(\mu^{-1}\wt \mu \ws)^{q_i} (i) = i$, and 
which defines a $q_i$-cycle of $\mu^{-1}\wt \mu \ws$. (The case $q_i=1$ has been dealt above.)
 In the product \eqref{cycles}, 
we collect all contraction $\delta$'s involved in the cycle starting at some fixed $i$   
\bea
\prod_{l=1}^{q_i}
\delta^{a_{(\mu^{-1}\wt \mu \ws)^{l-1}(i)} }_{  a_{(\mu^{-1}\wt \mu \ws)^l(i)}   }  \,. 
\label{breakcycle} 
\eea
Since this product is at arbitrary $i$, we have a companion and distinct product  of contraction  $\delta$'s that
starts at $\ws(i)$: $\prod_{l=1}^{q_i}
\delta^{a_{(\mu^{-1}\wt \mu \ws)^{l-1}( \ws (i))} }_{  a_{ (\mu^{-1}\wt \mu \ws)^l ( \ws (i))}   }$. 
Hence, we combine both products and multiply by one vertex $\delta$
\bea
\delta^{ a_{i}} _{ a_{\ws(i)}} 
\prod_{l=1}^{q_i}
\delta^{a_{(\mu^{-1}\wt \mu \ws)^{l-1}(i)} }_{  a_{(\mu^{-1}\wt \mu \ws)^l(i)}   }\, 
\delta^{a_{(\mu^{-1}\wt \mu \ws)^{l-1}( \ws (i))} }_{  a_{ (\mu^{-1}\wt \mu \ws)^l ( \ws (i))}   }
\label{breakcycle3} 
\eea
which evaluates to $N$ after performing the sum over the corresponding $a_j$'s. Again, the $q_i$-cycle is  counted once.  
It just remains to observe that the cycles, each defined by a subset of indices $a_j$, 
define partitions of the entire set of indices $a_i$ (once an index is used in a cycle it cannot appear in another one).   
Thus, the sum over $a_i$ factorizes along cycles and this complete the proof. 

\qed 

Note that there may be alternative ways of defining real tensor observables using pairings and without introducing the gauge redundancy. 
In any case, we could work in this setting, keeping track of the necessary information.

From Lemma \ref{lem1} applied to each color $i=1,2,3$, we finally come to 
\bea\label{2ptcorr}
\la O_{\s_1, \s_2, \s_3} (T)O_{\tau_1, \tau_2, \tau_3} (T)  \ra =  \sum_{\mu } N^{ \sum_{i=1}^3 \cy(\mu^{-1}\wt_i \mu \ws_i)}\,. 
\eea
The 1pt-correlator can be recovered from the above discussion. First, 
 the 1pt-correlator cannot be normal ordered. Introduce the Wick contraction  $\mu$ 
that belongs to $S_{2n}^*$ the subset defined by the pairings of $S_{2n}$ 
(a permutation pairing is made only of transpositions).  Then,  we obtain
\bea\label{1ptcorr}
\la O_{\s_1, \s_2, \s_3} (T)  \ra =  \sum_{\mu \in S_{2n}^* } 
 \sum_{a_{ij}} 
  \Big[\prod_{i=1}^{2n} \prod_{j=1}^{3} 
\delta^{a_{ij}}_{ a_{\ws_j(i)j} } \Big] 
  \Big[\prod_{i=1}^{2n}\prod_{j=1}^3 \delta^{a_{ij}}_{a_{\mu(i)j}}\Big]\,.
\eea
Next, we adapt Lemma \ref{lem1} to $\sum_{a_{i}} 
\Big[
\prod_{i=1}^{2n}
\delta^{a_{i}}_{a_{\ws(i)}}
\delta^{a_{i} }_{  a_{\ws \mu (i)}   }\Big]
 = N^{\cy(  \ws \mu)} $, 
 and then we obtain 
\bea\label{1ptcorr2}
\la O_{\s_1, \s_2, \s_3} (T)  \ra =  \sum_{\mu \in S_{2n}^*  } N^{ \sum_{i=1}^3 \cy(\mu \ws_i )}\,.
\eea

\noindent{\bf Representation theoretic base and orthogonality -}
We re-express the 2pt-function in order to make explicit some of its properties. 
Inserting 3 auxiliary permutations $\alpha_i  \in S_{2n} $,  the above sum \eqref{2ptcorr} reads as 
\be
\label{correl0}
\la O_{\s_1, \s_2, \s_3} (T)O_{\tau_1, \tau_2, \tau_3} (T)  \ra = 
\sum_{\mu  } \sum_{\alpha_i  } N^{ \sum_{i=1}^3 \cy (\alpha_i ) }
\prod_{i=1}^3 \delta (\mu^{-1}\wt_i \mu \ws_i \alpha_i) = N^{6n}
\sum_{\mu  }\prod_{i=1}^3 \delta (\mu^{-1}\wt_i \mu \ws_i  \Omega_i)  , 
\ee
where we introduced the central element $\Omega_i = \sum_{\alpha_i  \in S_{2n}}  N^{\cy(\alpha_i )-2n} \alpha_i $. 
The proof of that rests on the equality $\cy (\alpha_i^{-1}) = \cy (\alpha_i)$ and  that holds because
each cycle has an inverse, a cycle of the same length. 
Then, we can re-express \eqref{correl0} as 
\bea
\label{correl1}
&&
\la O_{\s_1, \s_2, \s_3} (T)O_{\tau_1, \tau_2, \tau_3} (T)  \ra 
\cr\cr 
&&
= 
 N^{6n}
\sum_{\mu  }\bdel [ (\mu^{-1})^{\otimes 3} (\wt_1 \otimes  \wt_2 \otimes \wt_3)
 \mu^{\otimes 3} (\ws_1 \otimes  \ws_2 \otimes \ws_3)
 (\Omega_1 \otimes  \Omega_2 \otimes \Omega_3)] \crcr
 &&
 = N^{6n}
\sum_{\mu  } \bdel [ (\wt_1 \otimes  \wt_2 \otimes \wt_3)
 \mu^{\otimes 3} (\ws_1 \otimes  \ws_2 \otimes \ws_3)
  (\mu^{-1})^{\otimes 3}
 (\Omega_1 \otimes  \Omega_2 \otimes \Omega_3) ]\,, 
\eea
where in the last equation we use the fact the $\Omega_i$ are central.  
We introduce the representation  theoretic element by pairing  a base element $ Q^{R_1, R_2, R_3, \tau}$ \eqref{unbasis} 
and an observable  $O_{\s_1, \s_2, \s_3}$ as
\bea
&&
O^{R_1, R_2, R_3, \tau}  
= \sum_{\s_l} \bdel( Q^{R_1, R_2, R_3, \tau} \s_1^{-1} \otimes \s_2^{-1} \otimes \s_3^{-1} )  O_{\s_1, \s_2, \s_3} \crcr
&&
 = \kappa_{\vec{R}}   \Big[  \prod_{i=1}^3  \frac{\kappa_{R_i}}{2n!} \Big]  
\sum_{\s_l}
\sum_{p_l \, ,q_l} C^{R_1,R_2;R_3,\tau}_{q_1,q_2;q_3} 
\Big[ \prod_{i=1}^3 
B^{R_i; \,  tr}_{p_i} 
D^{R_i}_{p_i q_i}(\s_i)\Big]    O_{\s_1, \s_2, \s_3}
\eea
As a linear combination of observables, we can calculate their correlators: 
\bea
&&
\la  O^{R_1, R_2, R_3, \tau}   \, O^{R'_1, R'_2, R'_3, \tau'}   \ra
 =  N^{6n}\kappa_{\vec{R}}   \kappa_{\vec{R}'}    \Big[  \prod_{i=1}^3  \frac{\kappa_{R_i}}{2n!} \frac{\kappa_{R'_i}}{2n!} \Big]  \cr\cr
 &&
 \times 
\sum_{\mu  }\bdel \Bigg[ 
  \sum_{\s_l, \s'_l}
\sum_{p_l, q_l, p'_l ,q'_l} C^{R_1,R_2;R_3,\tau}_{q_1,q_2;q_3} C^{R'_1,R'_2;R'_3,\tau'}_{q'_1,q'_2;q'_3} 
\Big[ \prod_{i=1}^3 
B^{R_i; \,  tr}_{p_i} 
D^{R_i}_{p_i q_i}(\s_i)B^{R'_i; \,  tr}_{p'_i} 
D^{R'_i}_{p'_i q'_i}(\s'_i)\Big]    \crcr
&&
\times  (\ws'_1 \otimes  \ws'_2 \otimes \ws'_3)
 \mu^{\otimes 3} (\ws_1 \otimes  \ws_2 \otimes \ws_3)
  (\mu^{-1})^{\otimes 3}
 (\Omega_1 \otimes  \Omega_2 \otimes \Omega_3) \Bigg] \crcr
 && = 
   N^{6n}  \kappa_{\vec{R}}   \kappa_{\vec{R}'}   \Big[  \prod_{i=1}^3  \frac{\kappa_{R_i}}{2n!} \frac{\kappa_{R'_i}}{2n!} \Big] 
\sum_{\mu  }\bdel \Bigg[ 
\sum_{p_l, q_l, p'_l ,q'_l} C^{R_1,R_2;R_3,\tau}_{q_1,q_2;q_3} C^{R'_1,R'_2;R'_3,\tau'}_{q'_1,q'_2;q'_3} \crcr
&&
\Big[ \otimes_{i=1}^3 
B^{R'_i; \,  tr}_{p'_i} 
 \sum_{\s'_i}  (\s'_i)^{-1} \xi  
D^{R'_i}_{p'_i q'_i}(\s'_i) \s'_i \Big]   \mu^{\otimes 3}  
\Big[ \otimes_{i=1}^3 
  B^{R_i; \,  tr}_{p_i} 
 \sum_{\s_i}
  (\s_i)^{-1} \xi   
D^{R_i}_{p_i q_i}(\s_i) \s_i  \Big]  (\mu^{-1})^{\otimes 3}   \crcr
&& \times 
 (\Omega_1 \otimes  \Omega_2 \otimes \Omega_3) \Bigg] \, . 
\eea
Next, we introduce the operator $T_\xi : S_{2n} \to S_{2n}$ 
that acts on $S_{2n}$ as $T_\xi (\s) = \s^{-1} \xi \s  = \ws $ and extends by linearity on $\mC(S_{2n})$. The operator $T_\xi $ actually maps any permutation to 
a pairing. Its image in $\mC(S_{2n})$ is the vector subspace generated
by all pairings (more properties are derived in Appendix \ref{app:groualg}).  We re-express the above correlator as
 \bea
 &&
\la  O^{R_1, R_2, R_3, \tau}   \, O^{R'_1, R'_2, R'_3, \tau'}   \ra \crcr
&&
= 
 N^{6n}  \kappa_{\vec{R}}   \kappa_{\vec{R}'} 
\sum_{\mu  }\bdel \Bigg[ 
\sum_{p_l, q_l, p'_l ,q'_l} C^{R_1,R_2;R_3,\tau}_{q_1,q_2;q_3} C^{R'_1,R'_2;R'_3,\tau'}_{q'_1,q'_2;q'_3} \crcr
&& \times 
\Big[ \otimes_{i=1}^3 
B^{R'_i; \,  tr}_{p'_i} 
T_\xi Q^{R'_i}_{p'_i q'_i} \Big]   \mu^{\otimes 3}  
\Big[ \otimes_{i=1}^3 
  B^{R_i; \,  tr}_{p_i} 
  T_\xi 
Q^{R_i}_{p_i q_i}  \Big]  (\mu^{-1})^{\otimes 3}  
 (\Omega_1 \otimes  \Omega_2 \otimes \Omega_3) \Bigg] \crcr
 && 
 = 
 N^{6n}
\sum_{\mu  }\bdel \Bigg[ 
\Big[ T_\xi ^{\otimes 3 }
\sum_{p'_l ,q'_l}  C^{R'_1,R'_2;R'_3,\tau'}_{q'_1,q'_2;q'_3}
 \otimes_{i=1}^3 B^{R'_i; \,  tr}_{p'_i} 
Q^{R'_i}_{p'_i q'_i} \Big]   \mu^{\otimes 3}  \crcr
&& 
\Big[
 T_\xi ^{\otimes 3 }
\sum_{p_l, q_l} C^{R_1,R_2;R_3,\tau}_{q_1,q_2;q_3} 
\otimes_{i=1}^3 
  B^{R_i; \,  tr}_{p_i} 
Q^{R_i}_{p_i q_i}  \Big]  (\mu^{-1})^{\otimes 3}  
 (\Omega_1 \otimes  \Omega_2 \otimes \Omega_3) \Bigg] \crcr
 && 
  = 
 N^{6n}
\sum_{\mu  }\bdel \Bigg[ 
 (T_\xi ^{\otimes 3 } Q^{R'_1,R'_2,R'_3,\tau'})  \mu^{\otimes 3} 
 (T_\xi ^{\otimes 3 } Q^{R_1,R_2,R_3,\tau} ) (\mu^{-1})^{\otimes 3} 
 (\Omega_1 \otimes  \Omega_2 \otimes \Omega_3) \Bigg] \crcr
 &&
 = 
 N^{6n}(2n!)\,  \bdel \Big[ 
 (T_\xi ^{\otimes 3 } Q^{R'_1,R'_2,R'_3,\tau'} )
 (T_\xi ^{\otimes 3 } Q^{R_1,R_2,R_3,\tau} )
 (\Omega_1 \otimes  \Omega_2 \otimes \Omega_3) \Big]\,, 
 \label{correlRep0}
 \eea
 where we used the right diagonal invariance of the base $Q^{R'_1,R'_2,R'_3,\tau'} $ to achieve the last stage of the calculation.
 Hence, this correlator computed with the Gaussian measure of $O(N)$ tensor models in the normal order, regarded as 
 an inner product on the space of observables, corresponds to the
 group theoretic inner product of the algebra $\cK_3(2n)$ calculated on a product 
of the transformed base $T_\xi ^{\otimes 3 } Q^{R_1,R_2,R_3,\tau}$ with an insertion of the factor $\Omega_1 \otimes  \Omega_2 \otimes \Omega_3$. The action $T_\xi ^{\otimes 3 }$ on $Q^{R_1,R_2,R_3,\tau}$ reflects the fact that it is the triple $(\ws_1, \ws_2, \ws_3)$ which plays a major role for computing the cycles associated with Feynman amplitudes 
in this theory (meanwhile the triple $(\s_1, \s_2, \s_3)$ was associated with the class counting  of the double coset space and its resulting algebra).
In $U(N)$ models \cite{BenRamg}, there is a correspondence between Gaussian 2pt-correlators in normal order and
the  inner product on the algebra of observables but without the presence of the operator $T_\xi ^{\otimes 3 }$. 
The presence of $T_\xi ^{\otimes 3 }$ determines therefore a  feature proper to $O(N)$ tensor models.  

We can further evaluate the above inner product as in Appendix \ref{app:correlator}
and find: 
\bea\label{correlatorRepr}
&& 
\la  O^{R_1, R_2, R_3, \tau}   \, O^{R'_1, R'_2, R'_3, \tau'}   \ra  = 
\Big[  \prod_{i=1}^3 \delta_{R'_i R_i } \Big]\delta_{\tau'\tau}
F(R_1, R_2, R_3,\tau) \crcr
&&
F(R_1, R_2, R_3,\tau) =  \sum_{S_i,\tau_i}
  \Big[ \prod_{i=1}^3   \Dim_N(S_i) \Big]
\Big[\sum_{b_i,c_i,p_i}
D^{S_i}_{b_ic_i}(\xi) 
C^{S_i,S_i;R_i,\tau_i}_{b_i,c_i;p_i}
B^{R_i; \,  tr}_{p_i}
\Big] ^2
\eea
which expresses the orthogonality of the representation theoretic base $\{O^{R_1, R_2, R_3, \tau} \}$ 
(corresponding to  normal ordered Gaussian correlators) of $\cK_3(2n)$.  
Note also that the pairing between  base elements is a representation translation of the Gaussian integration.

\ 

\noindent{\bf Rank $d$ 2pt-correlator -}
We obtain the 2pt-correlator at rank $d$ in a  straightforward  manner 
from the above derivation. We generalize \eqref{observ} and \eqref{correl2}
by extending the product over $j$ up to $d\ge 3$ and considering a tensor $T_{a_{i1}a_{i2} \dots a_{id}}$. 
The calculations are direct: we get \eqref{2ptcorr} and \eqref{1ptcorr2} by changing 
the sum over $i$ running over the colored cycles up to $d$. 
Meanwhile, the orthogonality of the 2pt-function is a property specific to the rank 3
and cannot be reproduced easily at any rank.

\

 \section{On $Sp(2N)$ tensor invariants}
 \label{spn}

We provide a few remarks on the counting of real $Sp(2N)$ tensor invariants. 
Carrozza and Pozsgay recently addressed  symplectic complex tensor models 
 in the context of tensor-like SYK models \cite{Carrozza:2018psc}. The authors focused on the complex group $U(N) \cap Sp(2N, \mathbb{C})$ (its quantum mechanical tensor model admits a large $N $
 expansion and shares similar properties of the SYK model) and, 
 at the combinatorial level, on the  improvement of the numerical computations of the number of its singlets in rank 3. We could ask, in the same vein as discussed above using symmetric group formulae, how to enumerate real symplectic invariants in the pure tensor model setting, i.e. with no spacetime attached to the tensor. 
We stress that, unlike in \cite{Carrozza:2018psc}, we are interested in real and Bosonic fields
and address in the following the symplectic group itself $Sp(2N,\mathbb{R}) = Sp(2N)$ and
its - symplectic - invariants in any rank. 
We show below that they follow an enumeration principle with the same
diagrammatics of that of the $O(N)$ invariants but  some changes occur at the level of the coset equivalence relation. Interestingly in this $Sp(2N)$ setting, the ``virtual'' vertices $v^i_j$, in Figure \ref{fig:coloredgraphs}, find an interpretation: their correspond precisely  to 
symplectic matrix $J$ insertions in the $Sp(2N)$ invariants. 

Let us recall the usual notation and introduce the real $2N\times 2N$ symplectic matrix $J$ which 
writes in blocks
\bea
 J  =\left(\begin{array}{cc}
0 & I_N \cr
- I_N & 0
\end{array}
\right)\,, 
\qquad J^2 = - I_{2N} \,, 
\eea
where $I_N$, for all $N$, is the identity matrix of $M_N(\mathbb{R})$. 
A matrix $K \in Sp(2N)$ obeys $K J K^{T} = J, $ and  $K^T J K = J$. 

A rank $d$ real tensor $T$, with components  $T_{p_1,\dots, p_d}$, $p_j=1,\dots, 2N$, transforms under the fundamental 
representation of $\otimes_{a=1}^d Sp(2N_a)$ for fixed $N_a$, if each group $Sp(2N_a)$ 
acts on the index $p_a$ such that the transformed tensor satisfies: 
\be
T^{K}_{q_1,\dots, q_d} = 
\sum_{p_1,\dots p_d} 
K^{(1)}_{q_1p_1} \dots K^{(d)}_{q_1p_1} \;
T_{p_1,\dots  , p_d} \,, 
\ee
where $K^{(a)} \in Sp(2N_a)$, $a=1,\dots, d$. 

Observables in $Sp(2N)$ tensor models are the contractions of an even number of tensors $T$. 
They are invariant under $\otimes_{a=1}^d Sp(2N_a)$  transformations and 
we call them $Sp(2N)$ invariants. 

In understood notation, we define a new trace on two rank $d$ tensors as
\be\label{traJ}
\Tr (T\,J^d \,T) = \sum_{p_i,q_i}
J^{(1)}_{p_1q_1} 
J^{(2)}_{p_2q_2} \dots J^{(d)}_{p_dq_d}\;\,
T_{p_1,\dots,p_d} T_{q_1,\dots,q_d} \,. 
\ee
Thus, the tensor indices  that are contracted 
couple with $J$.  This is the generalization of the symplectic
form over matrices which is defined as $\omega_J (M, W) = \tr (M^T J W),$ 
and that is invariant under symplectomorphisms. 

We check that $\Tr(T\,J^d \,T)$ is invariant under symplectic transformations: 
\bea
&&
\Tr (T^K\,J^d \,T^K) = 
\crcr
&& 
 \sum_{r_i,s_i}\;\; 
\sum_{p_i,q_i}
\big(K_{p_1,r_1} K_{q_1,s_1} J^{(1)}_{p_1q_1} \big)
\dots 
\big( K_{p_d,r_d}  K_{q_d,s_d}
 J^{(d)}_{p_dq_d}\big)
 T_{r_1,\dots, r_d} T_{s_1,\dots, s_d} 
 = \Tr(T J^d T)\,. 
\eea

Now, we extend the trace \eqref{traJ} to arbitrary number of tensors. 
Still the contraction obtained is an $Sp(2N)$ invariant. 
We can easily observe that the $Sp(2N)$ invariants can be viewed once again 
in terms of `$d$-regular colored graphs with a decoration on each edge. The decoration seals the
symplectic matrix $J$ on each pair of contracted tensor indices. 
Therefore, $J$  can be represented by a new vertex on each edge
which precisely plays the same role of a black vertex $v^i_j$ in  Figure \ref{fig:coloredgraphs}. 

The counting of $Sp(2N)$ invariants is more subtle than that of $O(N)$ invariants. Indeed, for simplicity, let
us consider in rank $3$ (generalizing the following argument at any rank $d$ is
straightforward), $2n$ tensors and count the possible triples $( \s_1 , \s_2 , \s_3 ) \in  S_{2n} \times S_{2n}  \times S_{2n}   $
subjected to the following invariance: 
\bea
( \s_1 , \s_2 , \s_3 ) \sim ( \gamma_1 \s_1 \gamma , \gamma_2 \s_2 \gamma , \gamma_3 \s_3 \gamma ) 
\eea
where, on the right, we have the ordinary diagonal action of $ \diag ( S_{2n}  )$ 
on the triple. Meanwhile, on the left, the $\gamma_i$ belong to an identical subgroup $G_i = G'$ but
that is not any more $S_n[S_2]$.  Switching the half-edges of the vertices $v^i_j$ produces a sign. This hints the fact that we should 
 switch to the group algebra $\mathbb{C}( S_{2n}) \times \mathbb{C}(S_{2n} ) \times \mathbb{C}( S_{2n} ) $
to perform the coset. At this point, note that nothing excludes
that the number of $Sp(2N)$ invariants matches the number of orthogonal invariants. 
Such interesting questions require much more work and is left  for future investigations. 

Let us make a final small remark. 
At this moment, we can give a precision about the complete graph, namely $K_4$,  that 
is identically vanishing in the complex Bosonic model with $U(N) \cap Sp(2N, \mathbb{C})$ invariance, as shown in  \cite{Carrozza:2018psc}. 
 In the present setting, we can show that it remains a nontrivial rank 3 symplectic invariant. We have developed
a code proving this fact for $Sp(2N=4)$. See the last code of Appendix \ref{app:mathsage}.  Of course here
$2N=4$ is not large and rather fixed, and one may question its physical interest. However, it is encouraging to see that it is not identically zero as its counterpart described above. Such a $K_4$ invariant
plays a central role in the study of the large $N$ and IR spectrum of the so-called ladder operators in the tensor-like SYK models. 
Hence, working with real Bosonic fields but with real $Sp(2N)$ invariance might become an important axis of research
in that direction. 

\section{Conclusion}
\label{concl}

This paper paves the way to a new formulation of real tensor models, 
their observables and correlators in terms of symmetric groups and its representation theory. 
The formulation is particularly convenient for implementing heavy computations using software resources, 
thus, leading to a gain of confidence in the computational process.  Furthermore, with its multiple facets, the  formalism elaborated here
may shed a different light on the same results since it bridges theories, combinatorics, TFT and physics through observables and correlators, which from the outset may look rather different. 

We have enumerated $O(N)$ or rank $d$ real tensor invariants as $d$-regular colored graphs
using a permutation group formalism. These invariants define the points of a double coset of $S_{2n}^{\times d}$. 
We use Mathematica and Sage codes to generate the
sequences associated with the number of these invariants from their generating functions. 
The sequences obtained at $d\ge 4$ are new according to the OEIS. 
Translated in the TFT$_2$ formulation, the same counting delivers the number of covers
of gluing of cylinders with defects. Such covers have been also observed while counting Feynman graphs
of scalar field theory \cite{FeynCount} and relate to a string theory on cylinders. 
Thus, there should be an equivalent way of describing tensor observables in  purely string theory language. Moreover, this link with covers must be made precise: covers in 2D 
are related to holomorphic maps and may, in return, give a geometry  to the space of orthogonal invariants. 
This point fully deserves further investigation.

Another piece of information reveals itself with the representation theoretic formulation of the counting: 
the number of orthogonal invariants is a sum of constrained Kronecker coefficients.  
The Kronecker coefficient is a core object in Computational Complexity theory: either finding a combinatorial 
rule describing it (finding which combinatorial objects  it counts), or its vanishing property or otherwise 
remain under active investigation (see  references in \cite{iken1, Blasiak}).  It concentrates a lot of research efforts 
since one expects that, roughly speaking,  an understanding that object could lead to a separation of complexity classes P vs NP. 
In our present work (and in a similar way in \cite{BenSamj}), we show that the number of tensor model observables - represented 
by colored graphs and thus combinatorial structures - links to a sum of Kronecker coefficients (in \cite{BenSamj}, it is a sum of square of these coefficients). It remains of course the question: how this would help  with one of the famous problems stated above? 
Perhaps a refined counting of colored graphs (endowed with specific properties) could boil down the sum to a single Kronecker element.
Such a study could bring some progress in the field.

The equivalence classes associated with the colored graphs are mapped in the tensor product 
of the group algebra $\mathbb{C}[S_{2n}]^{\otimes d}$. They form the base vectors of a  
subspace, namely $\cK_d(2n)$, that is in fact a semi-simple algebra. We call it a double coset algebra. 
Note also that, as element of an the algebra, $d$-regular colored graphs multiply in a specific way, and yield back 
a combination of $d$-regular  colored graphs. 
In rank 3, we have found ``natural'' representation theoretic base, $\{Q^{R,S,T, \tau}\}$, of $\cK_3(2n)$,  that means 
invariant and orthonormal. Unlike the unitary case \cite{BenSamj}, this base  decomposes in
blocks the algebra but does not provide its Wedderburn-Artin (WA) decomposition in matrix subalgebras. This brings other questions:  
 in which base the WA decomposition is made explicit? Is there a simple enough combination  
 starting from $Q^{R,S,T, \tau}$ that produces that WA decomposition?  A starting point of that analysis 
 might be given by the work by Bremner \cite{bremner} that constructs the WA base of a finite dimensional unital algebra over rationals.   Finally, is there a way 
 to understand why the sum of constrained Kronecker coefficients is actually a sum of squares
 (each of which  is the dimension of a matrix subalgebra entering in the WA decomposition)? 
 Such points deserve future  clarifications. 
 
 We  also addressed normal ordered Gaussian 2pt correlators in this work and show that, they formulate completely as 
a function of the size $N$ of the tensor indices and permutation cycles.  We generate an orthogonal representation 
base from  these 2pt correlators. This result is similar to what is observed in the unitary case, 
 with the following distinction: there is an operator acting on the triple defining the observables. 
We show that computing Gaussian correlators in representation theory space is actually computing
an inner product.  Finally, we briefly sketch the main feature of $Sp(2N)$ invariants: 
although they obey the same diagrammatics of the $O(N)$ invariants, they satisfy a different rule
concerning their equivalence classes. Thus, for the symplectic group and its invariants,
the story could be radically different from the orthogonal case and will require need more work.

\section*{Acknowledgments}
JBG  thankfully acknowledges discussions with Christophe Tollu, Sanjaye Ramgoolam and Pablo Diaz. 
RCA was supported by ISF Grant 1050/16.
JBG thanks the Laboratoire de Physique Th\'eorique d'Orsay for its hospitality when part of this
work was performed. 
JBG acknowledges a visiting fellowship of Perimeter Institute for Theoretical Physics.
This work is  supported by Perimeter Institute for Theoretical Physics. Research at Perimeter Institute is supported by the Government of Canada through Industry Canada and by the Province of Ontario through the Ministry of Research and Innovation.

\section*{ Appendix}

\appendix

\renewcommand{\theequation}{\Alph{section}.\arabic{equation}}
\setcounter{equation}{0}

\section{Symmetric group and its representation theory}
\label{app:SRT}

This appendix gathers useful identities and notations about the
symmetric group $S_n$ and its representation theory. 
The presentation here is a summary of Appendix A, withdrawn from \cite{BenSamj},
and the textbook by Hammermesh \cite{Hammermesh}. 

\subsection{Representation theory of the symmetric group} \label{app:reptheor}

Let $n$ be a positive integer and $S_n$, the group of permutation of $n$ elements. 
The Young diagrams or partitions $R$ of $n$, denoted  $ R \vdash n$, label the irreducible representations (irreps) of $S_n$. Consider $V_R$ a space of dimension $d(R)$ (that will 
be made explicit below). 
An irreps $\varrho_{R}: S_n \to {\rm End}(V_{R})$
is given by a matrix $D^{R}$ with entries 
$\varrho_{R} (\s) |R, i \rangle  =\sum_{l=1}^{d(R)} D^{R}_{li}(\s)|R, l \rangle$ with $\s\in S_n$
and with $ | R , i \ra $, $ i=1,\dots, d(R)$, an orthogonal base of states for $V_R$
(this base obeys $ \la R , j | R , i \ra = \delta_{ij}$).

We write in short $\varrho_R (\s) = \s$
and then $\langle R,j |\s|R, i \rangle = D^{R}_{ji}(\s)$. 
It is common to assimilate the irreducible representation $\varrho_{R} $ and the
carrier space $V_R$ with their label $R$. 

From the commuting action of the unitary group $U(N)$ and $S_{n}$ on 
a tensor product space $V^{\otimes n}$, the  Schur-Weyl duality teaches us
that we associate an irreps $R$ of $S_n$ with an irreps of $U(N)$, provided
$N$ bounds the length $l(R)$ of the first column of $R$, in symbol $l(R)\leq N$.  

Let us denote $d(R)$ the dimension of $R$ and 
$\Dim_N(R)$ the dimension of an irreps of $U(N)$, then those are given by  
\be\label{dims}
d(R)   =  n! / h(R)  \,,   \qquad  
\Dim_N(R) = f_N(R)/ h(R) \,,
\ee 
where $h(R)$ is the product of the hook lengths and $f_N(R)$ is the products of  box weights 
given by 
$h(R) = \prod_{i,j}(c_j-j +r_i-i+1)$ and $f_N(R)= \prod_{i,j}(N-i+j)$;
 the pairs $(i,j)$ label the boxes of the Young diagram with $ i$ the row label and $j$ is the column label. The $i$'th row length is $r_i$ and  $c_j$ is the column length of the $ j$'th column. 
  
We now restrict to real representations and so $D^R_{ij}(\s)$ must be real matrices. The matrix satisfies the following properties: 
\bea
&&
\sum_{i} D^R_{ a i} ( \s) 
 D^R_{ ib  } ( \s' ) = D^R_{  ab  } ( \s \s') \,, 
 \qquad 
D^R_{ab}(id) = \delta_{ab} \,, \qquad 
D^{R}_{ij}(\s^{-1}) = D^{R}_{ji}(\s)\,,
\label{ddinv} \\ 
&& 
\label{ortho}
\sum_{\s \in S_n} D^R_{ij}(\s)  D^S_{kl}(\s)
 = \frac{n!}{d(R)} \,\delta_{RS}\,\delta_{ik}\delta_{jl} \quad \text{(orthogonality)}\,.
\eea

The character of a given irreps $R$ is simply the trace of $D^R(\s)$, $\chi^R(\s)=\Tr (D^R(\s))= \sum_iD^R_{ii}(\s)$. The Kronecker delta  $\delta(\s) $ of the symmetric group (defined to be equal 1 when $\s= id$ and 0 otherwise) decomposes as
$\delta(\s) = \sum_{R \vdash n} \frac{d(R)}{n!} \, \chi^R(\s) $. 

The following identities are easily proved using the orthogonality relations of the 
representation matrices: 
\bea
&& 
\sum_{\g \in S_n} \delta( \g \s \g^{-1} \tau^{-1})  = \sum_{R \vdash n} \chi^R(\s)\chi^R(\tau) 
\,,\qquad \sum_{ \s \in S_n}  \chi^R(\s) \chi^S(\s) = n! \, \delta_{RS} \;
\text{(orthogonality)}
\label{delgsg0} \crcr
&&\\ 
&& 
\sum_{\g\in S_n} \chi^R(A\g B\g^{-1} ) 
= \frac{n!}{d(R)} \chi^{R}(A)\chi^{R}(B) \quad  \stackrel{\text{If $B$ is a central element}}{=}\quad  n! \, \chi^R(AB)
 \label{chiAB0}
\eea
 Also a useful identity expresses as
 \be\label{chiN}
{ 1 \over n! } \sum_{ \s}   \chi^R ( \s) N^{ \cy( \s ) } = \Dim_N (R) \;,
\qquad 
\sum_{\s \in S_n} D^R_{ij}(\s) N^{\cy(\s)} = \delta_{ij} f_N(R) \,,
 \ee
 where  $\cy(\s)$ is the number of cycles of $\s$. 

Defining the central element $ \Omega \in \mC ( S_n)$,
as 
$\Omega = \sum_{ \sigma \in S_n } N^{n - \cy( \sigma ) } \sigma$, 
the first relation in \eqref{chiN} can be also written as 
\be
{ N^n\over n! }  \chi^R ( \Omega ) = \Dim_N ( R ) \,.
\label{chiN2}
\ee

\subsection{Clebsch-Gordan coefficients }
\label{app:cgc}

Consider two carrier spaces $V_{R_1} $ and $ V_{R_2}  $ of two irreps of $S_n$ labeled by two Young diagrams $R_1$,
and $R_2 $, respectively.
The tensor product 
representation  $ V_{R_1}  \otimes V_{R_2}  $ can be decomposed into a direct sum of 
irreps $V_{R_3} $ with multiplicities
\bea 
V_{ R_1 } \otimes V_{R_2}  = \bigoplus_{ R_3 \vdash n } V_{R_3} \otimes V_{R_3}^{ \rm m } \,. 
\eea
The tensor product space  is spanned by a tensor product of the base $| R_1, i_1\ra\otimes |R_2,i_2\ra = :|R_1,i_1;R_2,i_2 \ra  $. On the right hand side,  the direct sum 
corresponds to a base set  $ | R_3 , i_3 , \tau_{R_3} \ra $. The label $i_3$ runs over states of $R_3$, and $ \tau_{R_3}$, the so-called multiplicity, runs over an orthogonal base in the multiplicity space $V_{R_3}^{\rm m}$. 

The Clebsch-Gordan coefficients are the branching coefficients between these bases: 
\be 
C^{R_1,R_2;\, R_3 ,\,\tau_{R_3} }_{\, i_1,i_2;\, i_3} :=    \la R_1,i_1; R_2,i_2 | R_3, \tau_{R_3}, i_3 \ra  = \la  R_3 , \tau_{R_3}, i_3 | R_1,i_1; R_2, i_2 \ra 
\ee
Note that they are real.

The following relations are detailed in Appendix A.2 in \cite{BenSamj}: 
\bea
&&
\sum_{j_1,j_2}
D^{R_1}_{i_1 j_1}(\g)D^{R_2}_{i_2 j_2}(\g)C^{R_1,R_2; \, R_3,\,\tau}_{\, j_1,j_2; \, j_3}
 =\sum_{i_3 }  C^{R_1,R_2;\, R_3,\,\tau}_{\, i_1,i_2; \, i_3} \, D^{R}_{i_3 j_3}(\g)   \,;
\label{ddc=cd} \\
&&
\sum_{i_1,i_2} 
C^{R_1,R_2; \, R_3 ,\,\tau}_{\, i_1,i_2; \,i_3} C^{R_1,R_2; \, R_3',\,\tau'}_{\, i_1,i_2; \,j_3}  
 = \delta_{R_3 R'_3 }\, \delta_{\tau\tau'} \, \delta_{i_3j_3}   \,;
\label{cc=del} \\
&&
\sum_{R_3 ,i_3 , \tau } 
C^{R_1,R_2; \, R_3 ,\,\tau}_{\, i_1,i_2; \, i_3 } C^{R_1,R_2; \, R_3 ,\,\tau}_{\, j_1,j_2; \,i_3 }  
 = \delta_{i_1j_1}\, \delta_{i_2j_2}   \,;
\label{cc=del2}  \\
&&
\sum_{R_3 ,\tau;\, i_3,j_3} 
C^{R_1,R_2; \, R_3,\,\tau}_{\, i_1,i_2; \,i_3} D^{R_3}_{i_3j_3}(\g) C^{R_1,R_2; \, R_3,\,\tau}_{\, j_1,j_2; \,j_3} 
=  D^{R_1}_{i_1j_1}(\g)D^{R_2}_{i_2j_2}(\g)    \,;
\label{dsd4}\\
&& 
\sum_{j_1,j_2,j_3}
D^{R_1}_{i_1 j_1}(\g)D^{R_2}_{i_2 j_2}(\g) D^{R_3}_{i_3 j_3}(\g) C^{R_1,R_2; \, R_3,\,\tau}_{\, j_1,j_2; \,j_3} 
 =   C^{R_1,R_2;\, R_3,\,\tau}_{\, i_1,i_2; i_3}   \,;
\label{dddc=c} 
\\
&& \sum_{i_l,j_l}
C^{R_1,R_2;R_3,\tau_1}_{i_1,i_2;i_3}
C^{R_1,R_2;R_3,\tau_2}_{j_1,j_2;j_3}
D^{R_1}_{i_1j_1}(\g_1\s_1\g_2)D^{R_2}_{i_2j_2}(\g_1\s_{2}\g_2)D^{R_3}_{i_3j_3}(\g_1\s_{3}\g_2)
 = \cr\cr
&&
\sum_{i_l,j_l}
C^{R_1,R_2;R_3,\tau_1}_{i_1,i_2;i_3}
C^{R_1,R_2;R_3,\tau_2}_{j_1,j_2;j_3}
 D^{R_1}_{i_1j_1}(\s_{1}) 
 D^{R_2}_{i_2j_2}(\s_{2})
 D^{R_3}_{i_3j_3}(\s_{3})   \,;
 \label{lem:2C3D}
\\
&&
\sum_{\s\in S_n} D^{R_1}_{i_1j_1}(\s)D^{R_2}_{i_2j_2}(\s)D^{R_3}_{i_3j_3}(\s)
 = \frac{n!}{d(R_3)}\sum_{\tau}
C^{R_1,R_2;R_3,\tau}_{i_1,i_2;i_3}
C^{R_1,R_2;R_3,\tau}_{j_1,j_2;j_3} \,. 
\label{lem2:DDD=CC}
\eea

Furthermore, we can generalize the second relation \eqref{chiN} as follows: given two permutations $A$ 
and $B$, we have 
\bea
&&
\sum_{\s \in S_n} D^R_{ij}(\s) N^{\cy(\s^{-1} A \s B)} =
\sum_{\g ,\s \in S_n} D^R_{ij}(\s) \delta(\g^{-1}\s^{-1} A \s B) N^{\cy(\g)} \crcr
&&
 = 
\sum_{S,a} \frac{d(S)}{n!} \sum_{\g ,\s} D^R_{ij}(\s) D^S_{aa}(\g^{-1}\s^{-1} A \s B) N^{\cy(\g)} 
\label{sumD} \\
&&
 = 
\sum_{S,a} \frac{d(S)}{n!} \sum_{m,n,o,p} \Big[\sum_{\g} D^S_{ma}(\g)  N^{\cy(\g)}\Big]
\Big[\sum_{\s } 
D^S_{nm}(\s) 
D^S_{op}(\s)
D^R_{ij}(\s)  
\Big]
D^S_{no}(A) 
D^S_{pa}(B) \,,
\nonumber
\eea
with the property $\cy(\g) = \cy(\g^{-1})$. We now use \eqref{chiN} and \eqref{lem2:DDD=CC}
to write
\bea
&&
\sum_{\s \in S_n} D^R_{ij}(\s) N^{\cy(\s^{-1} A \s B)} = \crcr
&&
\sum_{S,a} \frac{d(S)}{n!} \sum_{m,n,o,p}  \delta_{ma} f_N(S) 
\Big( \frac{n!}{d(R)} \sum_{\tau}  C^{S,S; R, \tau}_{n,o;i} C^{S,S; R, \tau}_{m,p;j}\Big) 
D^S_{no}(A) 
D^S_{pa}(B)  \crcr
&&
=
\sum_{S, \tau} \frac{d(S)}{d(R)}  f_N(S)
\Big(  \sum_{n,o}  C^{S,S; R, \tau}_{n,o;i}  D^S_{no}(A)  \Big) 
\Big( \sum_{a,p}  C^{S,S; R, \tau}_{a,p;j} D^S_{pa}(B) \Big) \,.
\label{sumDN}
\eea

\subsection{Base of the group algebra $\mC ( S_n)$}
\label{app:groualg}

The matrix base of the group algebra  $ \mC ( S_n )$ is defined by the elements  
\be
Q^R_{ ij}   =   { \kappa_R \over n! } \sum_{ \sigma \in S_n } D^R_{ ij} ( \sigma ) \sigma  \, ,
\ee
where the constant $\kappa_R^2 = n!d(R)$  is a fixed by a normalization. The base set  $\{Q^R_{ij}\}$ is 
of cardinality  $\sum_{ R \vdash n } (d(R))^2 = n! $. 
The elements $Q^R_{ij}$ form a representation theoretic Fourier base for $ \mC ( S_n)$. 

The left and right multiplication by group elements on $Q^R_{ij}$  expand as 
\be \label{tauq}
\tau \,  Q^R_{ ij} = 
\sum_{l} D^R_{ li} ( \tau ) \,  Q^R_{ l j} \,, \qquad  
 Q^R_{ ij} \, \tau  =    \sum_{l} Q^R_{ i  l} \, D^R_{ j l } ( \tau )  \, .
\ee

Using the definition of the base and \eqref{tauq}, one gets
\beq \begin{aligned}
Q^R_{ij} Q^{R'}_{kl} &= 
\frac{\kappa_R \kappa_{R'}}{(n!)^2} \sum_{\sigma \in S_{n}} \sum_{\tau \in S_{n}} D^R_{ij}(\sigma) \sigma D^{R'}_{kl}(\tau)\tau 
= \frac{\kappa_R}{n!}  \sum_{\sigma \in S_{n}} D^R_{ij}(\sigma) \sigma Q^{R'}_{kl}\\
& = \frac{\kappa_R}{n!} \sum_{\sigma \in S_{n}} D^R_{ij}(\sigma) 
\sum_{m} D^{R'}_{mk}(\sigma) Q^{R'}_{ml} 
= \frac{\kappa_R}{n!} \sum_m \frac{n!}{d(R)} \delta_{RR'}\delta_{im} \delta_{jk} Q^{R'}_{ml} \\
& = \frac{\kappa_R}{d(R)}\delta_{RR'}\delta_{jk} Q^{R'}_{il}.
\end{aligned} 
\eeq

We consider the Kronecker $\delta$ on $S_n$, and
extend it (by linearity) as a pairing denoted again $\delta$ on $ \mC ( S_n )$, 
and then once again extend the result to $ \mC ( S_n ) ^{\otimes d}$, $d >1 $, such that 
\bea
\label{bdel}
\bdel(\s_1 \otimes \dots \otimes \s_d; 
\s_1' \otimes \dots \otimes \s_d') =\delta(\s_1\s_1'^{-1})\dots
\delta(\s_d^{-1}\s_d'^{-1})\,. 
\eea
Calculating the inner product $\delta(Q^{ R}_{ i j };Q^{ R'}_{ i' j' }) $, we obtain
\bea
\delta(Q^{ R}_{ i j };Q^{ R'}_{ i' j' }) 
 =\frac{\kappa_R^2}{n!d(R)}\delta_{RR'}\delta_{ii'}\delta_{jj'} 
= \delta_{RR'}\delta_{ii'}\delta_{jj'}   \,. 
\eea
 Then,  for multiple tensor factors, we obtain
\be
\bdel(
Q^{ R_1}_{ i_1 j_1 } \otimes \dots \otimes Q^{ R_d}_{ i_d  j_d }\,;\,
Q^{ R'_1}_{ i'_1 j'_1 } \otimes \dots \otimes Q^{ R'_d}_{ i'_d  j'_d })
 = 
\delta_{R_1R_1'}\delta_{i_1i_1'}\delta_{j_1j_1'}
\dots
\delta_{R_dR'_d}\delta_{i_di'_d}\delta_{j_dj'_d} \,. 
\label{pairing}
\ee
Hence, the base $\{Q^{ R_1}_{ i_1 j_1 } \otimes \dots \otimes Q^{ R_d}_{ i_d  j_d }\}$ is an 
Fourier theoretic orthonormal  base 
for $ \mC ( S_n )^{\otimes d}$. 

In the text, we focus on $S_{2n}$ and we introduce the operator  $T_\xi : S_{2n} \to S_{2n}$ 
that acts on $S_{2n}$ as $T_\xi (\s) = \s^{-1} \xi \s  $. In a natural way, $T_\xi $ extends by linearity on $\mC(S_{2n})$. Then, without any possible confusion with the tensor notation $T$ itself, 
$T_\xi \in {\rm End}(\mC(S_{2n}))$ is the image of a mapping $T: S_{2n}\to {\rm End}(\mC(S_{2n}))$, such that $\xi \mapsto T_\xi$. We then extend $T$ by linearity over
$T: \mC(S_{2n})\to {\rm End}(\mC(S_{2n}))$, such that $\lambda \xi + \rho  \mapsto T_{\lambda \xi + \rho} =  \lambda T_{ \xi }+ T_{ \rho}$, $\lambda \in \mC$.

We are interested in the properties of the transformed base $T_\xi Q^{ R}_{ ij }$ which is nothing but the Fourier transformed of the pairing $ \s^{-1} \xi \s$. First, let us
see how they multiply: 
\bea
(T_\xi Q^{ R}_{ ij })\, (T_\xi Q^{ R'}_{ i'j'}) = \frac{\kappa_R\kappa_{R'}}{(2n!)^2}
\sum_{\sigma ,\rho \in S_{2n}} D^R_{ij}(\sigma) D^{R'}_{ i'j'}(\rho) \s^{-1} \xi \s \rho^{-1}\xi \rho \,.
\eea
Note that the group order is now $2n!$. 
Introduce a change of variable $\s \to\s \rho^{-1}$, and 
\bea
&&
(T_\xi Q^{ R}_{ ij })\, (T_\xi Q^{ R'}_{ i'j'}) = \frac{\kappa_R\kappa_{R'}}{(2n!)^2}
\sum_{\sigma ,\rho \in S_{2n}} \sum_{k}D^R_{ik}(\sigma)D^R_{kj}(\rho ) D^{R'}_{ i'j'}(\rho) \rho^{-1} \s^{-1} \xi \s \xi \rho \crcr
&& = 
 \frac{\kappa_{R'}}{(2n!)}
\sum_{\rho \in S_{2n}}  D^{R'}_{ i'j'}(\rho) \sum_{k}D^R_{kj}(\rho )
T_{(T_\xi Q^R_{ik})\xi}(\rho) \crcr
&&
= \frac{\kappa_{R'}}{(2n!)}
 \sum_{\rho \in S_{2n}} D^{R'}_{ i'j'}(\rho) T_{ \sum_{k}D^R_{kj}(\rho )  (T_\xi Q^R_{ik}) \xi}( \rho) \,. 
\eea
Thus, the product of the transformed base elements does not re-express easily  in 
terms of the transformed base elements. 
The left and right multiplications of fixed permutations on the elements $T_\xi Q^{ R}_{ ij }$,  counterparts of \eqref{tauq}, 
are given by: 
\bea
\tau(T_\xi Q^{ R}_{ ij }) = \sum_{a} \, (T_\xi Q^{ R}_{ i a}) D^R_{aj}(\tau)\tau  \,,
\qquad 
(T_\xi Q^{ R}_{ ij }) \tau = \sum_{a} \, (T_\xi Q^{ R}_{ i a}) D^R_{ja}(\tau)\tau  \,. 
\eea
The inner product of these elements expresses as:
\bea
\delta(T_\xi Q^{ R}_{ ij },\, T_\xi Q^{ R'}_{ i'j'}) = 
 \frac{\kappa_R\kappa_{R'}}{(2n!)^2}
\sum_{\sigma ,\rho \in S_{2n}} D^R_{ij}(\sigma)D^{R'}_{ i'j'}(\rho)\, \delta (T_\xi (\s),T_\xi (\rho))\,. 
\eea
This  is simply the Fourier transform of the delta $\delta (\s^{-1} \xi \s  \rho^{-1} \xi \rho)$
which  tells us that  the sole  terms remaining in this sum are those which define the same pairing.
A closer look shows that $\delta (\s^{-1} \xi \s  \rho^{-1} \xi \rho) = \delta ( \xi \s  \rho^{-1} \xi \rho \s^{-1})$. Then, this means that the elements that  contribute
to the sum are those $\s  \rho^{-1} $ that belong to the 
stabilizer of $\xi$, that is $\s  \rho^{-1} \in S_n[S_2]$. Hence, we change variable
as
$\s \to \bar \s = \s  \rho^{-1}$, rename again $\bar \s $ as $\s$ and then rewrite,
using the orthogonality of the representation matrices: 
\bea
&&
\delta(T_\xi Q^{ R}_{ ij },\, T_\xi Q^{ R'}_{ i'j'}) = 
 \frac{\kappa_R\kappa_{R'}}{(2n!)^2}
\sum_{\rho \in S_{2n}} \sum_{\s \in S_n[S_2]} 
D^R_{ij}(\sigma \rho)D^{R'}_{ i'j'}(\rho) \crcr
&&
= \frac{\kappa_R\kappa_{R'}}{(2n!)^2}
\sum_a \sum_{\s \in S_n[S_2]} 
 D^R_{ia}(\sigma)\sum_{\rho \in S_{2n}} D^R_{aj}( \rho) D^{R'}_{ i'j'}(\rho)
 \crcr
&&
= \delta_{RR'} \delta_{jj'}\frac{\kappa_R^2}{(2n!)^2}
\frac{2n!}{d(R)} 
\sum_a \sum_{\s \in S_n[S_2]} 
 D^R_{ia}(\sigma)  \delta_{ai'} \crcr
&&
= \delta_{RR'} \delta_{jj'}\sum_{\s \in S_n[S_2]} 
 D^R_{ii'}(\sigma) \,. 
 \label{deltaTT}
\eea
 In the text, we compute a formula for that sum in terms
 of branching coefficients,  see \eqref{DSS}. It turns out that the sum is
  nonvanishing only if the partition $R$ is  even, meaning that the length of each of its rows is  even. Hence, from
  the above relation, \eqref{deltaTT}, the  set of the transformed base elements 
  does not form an orthogonal system. 
 
 It is instructive to perform the same evaluation in an alternative way to discover
 new identities satisfied by the Clebsch-Gordan coefficients.
 Consider the expansion of the above inner product as follows: 
\bea
&&
\delta(T_\xi Q^{ R}_{ ij },\, T_\xi Q^{ R'}_{ i'j'}) = 
 \frac{\kappa_R\kappa_{R'}}{(2n!)^2}\sum_{S} \frac{d(S)}{2n!}
\sum_{\sigma ,\rho \in S_{n}} D^R_{ij}(\sigma)D^{R'}_{ i'j'}(\rho)
\chi^S (\s^{-1} \xi \s  \rho^{-1} \xi \rho)\crcr
&&
= \frac{\kappa_R\kappa_{R'}}{(2n!)^2}\sum_{S} \frac{d(S)}{2n!}
 \sum_{a,b,c,d,e,f} D^S_{bc} (\xi) D^S_{ef} (\xi)
\sum_{\sigma ,\rho } 
D^S_{ba}(\s)
D^S_{cd} (\s)
D^R_{ij} (\s)
D^S_{fa} (\rho)
D^S_{ed} (\rho)
D^{R'}_{ i'j'}(\rho)
\crcr
&&
= \frac{\kappa_R\kappa_{R'}}{(2n!)^2}\sum_{S} \frac{d(S)}{2n!}
 \sum_{b,c,e,f} D^S_{bc} (\xi) D^S_{ef} (\xi)
  \frac{ (2n!)^2 }{ d(R)d(R')} 
 \sum_{\tau,\tau'}
C^{S,S;R,\tau}_{b,c;i} C^{S,S;R',\tau'}_{f,e;i'}
\sum_{a,d}
 C^{S,S;R,\tau}_{a,d;j}
  C^{S,S;R',\tau'}_{a,d;j'}
\crcr
&&
= \frac{\kappa_R\kappa_{R'}}{ d(R)d(R')} 
\sum_{S} \frac{d(S)}{2n!}
 \sum_{b,c,e,f} D^S_{bc} (\xi) D^S_{ef} (\xi)
 \sum_{\tau,\tau'}
C^{S,S;R,\tau}_{b,c;i} C^{S,S;R',\tau'}_{f,e;i'}
\delta_{RR'}\delta_{\tau\tau'}\delta_{jj'}
  \crcr
&&
= \delta_{RR'} \delta_{jj'} \frac{\kappa_R^2}{d(R)^2} 
\sum_{S,\tau} \frac{d(S)}{2n!}
 \sum_{b,c,e,f} D^S_{bc} (\xi) D^S_{ef} (\xi)
C^{S,S;R,\tau}_{b,c;i} C^{S,S;R,\tau}_{f,e;i'}
  \crcr
&&  = 
 \delta_{RR'} \delta_{jj'} \frac{1}{d(R)} 
\sum_{S,\tau} d(S)
F(S,R, \tau;i) F(S,R,\tau;i')\, , 
\label{deltaTQTQ}
\eea
where, at some intermediate steps, we used successively \eqref{lem2:DDD=CC} and \eqref{cc=del}, and where $F(S,R,\tau;i)=  \sum_{b,c} D^S_{bc} (\xi)C^{S,S;R,\tau}_{b,c;i}$. 
Using $\sum_{\s\in S_{n}[S_2]} D^{R}_{ij}(\s) = 
(2^n n!) B^{ R; \,  tr}_{i} B^{R;\,  tr}_{j}$ (see \eqref{DSS}), we arrive to a new identity: 
\be
\sum_{S,\tau} d(S)
\big(\sum_{b,c} D^S_{bc} (\xi)C^{S,S;R,\tau}_{b,c;i}\big)
\big(\sum_{e,f} D^S_{ef} (\xi)C^{S,S;R,\tau}_{e,f;j}\big)  = 
 \frac{(2^n n!)}{d(R)}  B^{ R; \,  tr}_{i} B^{R;\,  tr}_{j} \,.
 \label{sumbb}
\ee
Note the similarity of the left-hand-side member with \eqref{sumDN} (adjusted for the symmetric group $S_{2n}$). 

There exist graphical ways of representing identities in representation 
theory in general. For the permutation group, Appendix A2 of \cite{BenSamj}
lists such graphical  representations for most of the identities given above. For instance, we use the graphical representation of the representation matrix $D^R_{ij}(\s)$ as $\mytikz{
		\node (g) at (0,0) [rectangle,draw] {$\s$};
		\node (i1) at (-0.4,0.2) {$i$};
	       \node (i2) at (0.4,0.2) {$j$};
		\draw (-0.5,0) -- (-0.26,0) ; 
		\draw (0.26,0) -- (0.5,0) ;  
	}$, the Clebsch-Gordan coefficient 
$C^{R_2 , R_2 ;R_3 , \tau}_{i_1,i_2; i_3 }$ represents as follows $
\mytikz{	
	\node (t) at (0,0) [circle,fill,inner sep=0.5mm,label=below:$\tau$] {};	
	\node (i1) at (-1.5,0.7) {$i_1$};
	\node (i2) at (-1.5,-0.8) {$i_2$};
	\node (m) at (1.5,0) {$ i_3 $};
	\draw [postaction={decorate}] (t) to node[above]{$R_3$} (m);		
	\draw [postaction={decorate}] (i1) to node[above]{$R_1$} (t);
	\draw [postaction={decorate}] (i2) to node[below]{$R_2$} (t);
}$ 
and the branching coefficient 
$B^{R; \,r,\nu_r}_{i;\, m_r}$ looks like 
$\mytikz{
\node (i) at (-1,0) {$i$};		
\node (n) at (0,0) [circle,draw,inner sep=0.5mm,label=below:$\nu_r$] {};
\node (l2) at (1.5,0) {$m_r$};
\draw [postaction={decorate}] (i) to node[above]{$R$} (n);
\draw [postaction={decorate}] (n) to node[above right]{$r$} (l2);
}$.
Then the convolution given by  \eqref{sumbb} translates as the factorization: 
\bea
\sum_{S,\tau} d(S)
	\mytikz{				
	\node (n) at (-2.2,0) [circle,fill, draw,inner sep=0.5mm,label=below:$\tau$] {};	
	\node (tn) at (2.3,0) [circle,fill, draw,inner sep=0.5mm,label=below:$\tau$] {};	
	\node (s11) at (-0.3,0) [rectangle,draw] {$\xi$};
	\node (s12) at (0.5,0) [rectangle,draw] {$\xi$};
	\node (i) at (-3.2,0) {$i$};				
	\node (j) at (3.2,0) {$j$};							
	\draw [postaction={decorate}] (s11) to [bend left=40] node[below]{$S$} (n);
	\draw [postaction={decorate}] (s11) to [bend right=40] node[above]{$S$} (n);		
	\draw [postaction={decorate}] (s12) to [bend left=40] node[above]{$S$} (tn);
	\draw [postaction={decorate}] (s12) to [bend right=40] node[below]{$S$} (tn);		
	\draw [] (s12) to [bend left=40] (tn);
	\draw [] (s12) to [bend right=40] (tn);
	\draw [postaction={decorate}] (n) to node[above]{$R$} (i);
	\draw [postaction={decorate}] (tn) to node[above]{$R$} (j);		
	}		
\; &= \;	 \frac{(2^n n!)}{d(R)} 
	\mytikz{
\node (i) at (-1,0) {$i$};		
\node (n) at (0,0) [circle,draw,inner sep=0.5mm,label=below:$1$] {};
\node (l2) at (1.5,0) {$0$};
\draw [postaction={decorate}] (i) to node[above]{$R$} (n);
\draw [postaction={decorate}] (n) to node[above right]{$tr$} (l2);
\node (j) at (-1,-1) {$j$};		
\node (m) at (0,-1) [circle,draw,inner sep=0.5mm,label=below:$1$] {};
\node (l3) at (1.5,-1) {$0$};
\draw [postaction={decorate}] (j) to node[above]{$R$} (m);
\draw [postaction={decorate}] (m) to node[above right]{$tr$} (l3);
}\,,
\eea
hence, a new identity satisfied by the Clebsch-Gordan of the symmetric group.

\subsection{2pt-correlator evaluation}
\label{app:correlator}

We prove in this part \eqref{correlatorRepr}. 
To proceed, we will make use of
\eqref{chiN},  \eqref{cc=del} and \eqref{lem2:DDD=CC}, or alternatively \eqref{sumDN}, of Appendix \ref{app:cgc}. 
Introducing  $k_{\vec{R}}  =\kappa_{\vec{R}}  
\frac{\kappa_{R_1}\kappa_{R_2}\kappa_{R_3}}{((2n)!)^3}$, then from 
\eqref{correlRep0}, we focus on the $\bdel$ function: 
\bea
&&
\bdel \Big[ 
 (T_\xi ^{\otimes 3 } Q^{R'_1,R'_2,R'_3,\tau'} )
 (T_\xi ^{\otimes 3 } Q^{R_1,R_2,R_3,\tau} )
 (\Omega_1 \otimes  \Omega_2 \otimes \Omega_3) \Big] 
\crcr
&&
 = k_{\vec{R}}k_{\vec{R}}'
\sum_{p_l, q_l, p'_l ,q'_l} C^{R_1,R_2;R_3,\tau}_{q_1,q_2;q_3} C^{R'_1,R'_2;R'_3,\tau'}_{q'_1,q'_2;q'_3} 
\Big[ \prod_{i=1}^3 B^{R'_i; \,  tr}_{p'_i} B^{R_i; \,  tr}_{p_i}  \Big] \crcr
&& \times 
 \sum_{\s'_i,\s_i}
 \sum_{\alpha_i} \Big[ \prod_{i=1}^3N^{\cy(\alpha_i)-2n}  \Big]
\Big[\prod_{i=1}^3  D^{R'_i}_{p'_i q'_i}(\s'_i) D^{R_i}_{p_i q_i}(\s_i)
\delta((\s'_i)^{-1} \xi   \s'_i   (\s_i)^{-1} \xi    \s_i \alpha_i ) \Big] \crcr 
&&
= k_{\vec{R}} k_{\vec{R}}'\sum_{p_l, q_l, p'_l ,q'_l} C^{R_1,R_2;R_3,\tau}_{q_1,q_2;q_3} C^{R'_1,R'_2;R'_3,\tau'}_{q'_1,q'_2;q'_3} 
\Big[ \prod_{i=1}^3 B^{R'_i; \,  tr}_{p'_i} B^{R_i; \,  tr}_{p_i}  \Big]
 \sum_{\s'_i,\s_i}
 \Big[\prod_{i=1}^3  D^{R'_i}_{p'_i q'_i}(\s'_i) D^{R_i}_{p_i q_i}(\s_i)\Big] \crcr
 &&
 \times 
  \sum_{S_i,a_i,g_i}
 \sum_{\alpha_i} \Big[ \prod_{i=1}^3N^{\cy(\alpha_i)-2n} D^{S_i}_{g_ia_i}(  \alpha_i ) \Big] \crcr
 && \times 
\sum_{b_i,c_i,d_i,e_i,f_i}
\prod_{i=1}^3
\frac{d(S_i)}{2n!}
D^{S_i}_{a_ib_i}((\s'_i)^{-1} )
D^{S_i}_{b_ic_i}(\xi)
D^{S_i}_{c_id_i}(\s'_i)
D^{S_i}_{d_ie_i}((\s_i)^{-1} )
D^{S_i}_{e_if_i}(\xi)
D^{S_i}_{f_ig_i}( \s_i )
\crcr 
&&
= k_{\vec{R}} k_{\vec{R}}'
 \frac{N^{-6n}}{(2n!)^3}\sum_{p_l, q_l, p'_l ,q'_l} C^{R_1,R_2;R_3,\tau}_{q_1,q_2;q_3} C^{R'_1,R'_2;R'_3,\tau'}_{q'_1,q'_2;q'_3} 
\Big[ \prod_{i=1}^3 B^{R'_i; \,  tr}_{p'_i} B^{R_i; \,  tr}_{p_i}  \Big]
 \sum_{S_i,a_i,g_i}
\Big[ \prod_{i=1}^3    \delta_{g_ia_i} f_N(S_i)d(S_i) \Big] \crcr
 && \times 
\sum_{b_i,c_i,d_i,e_i,f_i}
 \sum_{\s'_i,\s_i}
 \Big[\prod_{i=1}^3
D^{S_i}_{b_ia_i}(\s'_i )
D^{S_i}_{c_id_i}(\s'_i)
 D^{R'_i}_{p'_i q'_i}(\s'_i)
 D^{S_i}_{f_ig_i}( \s_i ) 
D^{S_i}_{e_id_i}(\s_i )
 D^{R_i}_{p_i q_i}(\s_i) \Big] 
\crcr 
&&
\times
\Big[ 
\prod_{i=1}^3
D^S_{b_ic_i}(\xi) D^S_{e_if_i}(\xi) \Big] \, .
\eea
It is the moment to use \eqref{lem2:DDD=CC} to integrate  the representation matrices
and get: 
\bea
&&
k_{\vec{R}} k_{\vec{R}}' \frac{N^{-6n}}{(2n!)^3}\sum_{p_l, q_l, p'_l ,q'_l} C^{R_1,R_2;R_3,\tau}_{q_1,q_2;q_3} C^{R'_1,R'_2;R'_3,\tau'}_{q'_1,q'_2;q'_3} 
\Big[ \prod_{i=1}^3 B^{R'_i; \,  tr}_{p'_i} B^{R_i; \,  tr}_{p_i}  \Big]
 \sum_{S_i}
\Big[ \prod_{i=1}^3  (2n!) \Dim_N(S_i) \Big] \crcr
 && \times 
\sum_{a_i,b_i,c_i,d_i,e_i,f_i}
\prod_{i=1}^3\Big[ \frac{2n!}{d(R_i)d(R_i')}
\sum_{\tau'_i,\tau_i}
C^{S_i,S_i;R'_i,\tau'_i}_{b_i,c_i;p'_i}
C^{S_i,S_i;R'_i,\tau'_i}_{a_i,d_i;q'_i}
C^{S_i,S_i;R_i,\tau_i}_{f_i,e_i;p_i}
C^{S_i,S_i;R_i,\tau_i}_{a_i,d_i;q_i} \Big] 
\crcr 
&&
\times
\Big[ 
\prod_{i=1}^3
D^{S_i}_{b_ic_i}(\xi) D^{S_i}_{e_if_i}(\xi) \Big] 
\crcr
&&
= k_{\vec{R}}^2 N^{-6n}\prod_{i=1}^3\Big[ \frac{2n!}{d(R_i)d(R_i')}\Big] \sum_{p_l, q_l, p'_l ,q'_l} C^{R_1,R_2;R_3,\tau}_{q_1,q_2;q_3} C^{R'_1,R'_2;R'_3,\tau'}_{q'_1,q'_2;q'_3} 
\Big[ \prod_{i=1}^3 B^{R'_i; \,  tr}_{p'_i} B^{R_i; \,  tr}_{p_i}  \Big]\crcr
 && \times 
 \sum_{S_i}
\Big[ \prod_{i=1}^3  \Dim_N(S_i) \Big] \crcr
 && \times 
\sum_{b_i,c_i,e_i,f_i}
\prod_{i=1}^3\Big[
\sum_{\tau'_i,\tau_i}
C^{S_i,S_i;R'_i,\tau'_i}_{b_i,c_i;p'_i}
C^{S_i,S_i;R_i,\tau_i}_{f_i,e_i;p_i}
\delta_{R'_i R_i }
\delta_{\tau'_i \tau_i}
\delta_{q'_i q_i} \Big] 
\Big[ 
\prod_{i=1}^3
D^{S_i}_{b_ic_i}(\xi) D^{S_i}_{e_if_i}(\xi) \Big]
 \crcr
&&
=   \frac{(2n!)^3 k_{\vec{R}}^2 N^{-6n} }{\prod_{i=1}^3 d(R_i)^2} \Big[  \prod_{i=1}^3 \delta_{R'_i R_i } \Big]
\sum_{p_l, q_l, p'_l } C^{R_1,R_2;R_3,\tau}_{q_1,q_2;q_3} C^{R_1,R_2;R_3,\tau'}_{q_1,q_2;q_3} 
\Big[ \prod_{i=1}^3 B^{R'_i; \,  tr}_{p'_i} B^{R_i; \,  tr}_{p_i}  \Big]
 \sum_{S_i}
\Big[ \prod_{i=1}^3   \Dim_N(S_i) \Big] \crcr
 && \times 
\sum_{b_i,c_i,e_i,f_i}
\prod_{i=1}^3\Big[
\sum_{\tau_i}
C^{S_i,S_i;R_i,\tau_i}_{b_i,c_i;p'_i}
C^{S_i,S_i;R_i,\tau_i}_{f_i,e_i;p_i} \Big] 
\Big[ 
\prod_{i=1}^3
D^{S_i}_{b_ic_i}(\xi) D^{S_i}_{e_if_i}(\xi) \Big]  
\crcr
&&
= \kappa_{\vec{R}}^2 N^{-6n} \Big[  \prod_{i=1}^3 \delta_{R'_i R_i } \Big]\delta_{\tau'\tau}
\sum_{ q_3 }  \delta_{q_3 q_3} 
\sum_{p_l, p'_l } 
\Big[ \prod_{i=1}^3 B^{R'_i; \,  tr}_{p'_i} B^{R_i; \,  tr}_{p_i}  \Big]
 \sum_{S_i}
\Big[ \prod_{i=1}^3  \Dim_N(S_i)\Big] \crcr
 && \times 
\sum_{b_i,c_i,e_i,f_i}
\prod_{i=1}^3\Big[
\sum_{\tau_i}
C^{S_i,S_i;R_i,\tau_i}_{b_i,c_i;p'_i}
C^{S_i,S_i;R_i,\tau_i}_{f_i,e_i;p_i} \Big] 
\Big[ 
\prod_{i=1}^3
D^{S_i}_{b_ic_i}(\xi) D^{S_i}_{e_if_i}(\xi) \Big] \crcr
&&
= \kappa_{\vec{R}}^2 d(R_3)N^{-6n}\Big[  \prod_{i=1}^3 \delta_{R'_i R_i } \Big]\delta_{\tau'\tau}
 \sum_{S_i,\tau_i}
  \Big[ \prod_{i=1}^3   \Dim_N(S_i) \Big] \crcr
 && \times 
\prod_{i=1}^3
\Big[\sum_{b_i,c_i,p'_i}
D^{S_i}_{b_ic_i}(\xi) 
C^{S_i,S_i;R_i,\tau_i}_{b_i,c_i;p'_i}
B^{R_i; \,  tr}_{p'_i}
\Big] 
\Big[
\sum_{p_i,e_i,f_i}
D^{S_i}_{e_if_i}(\xi) 
C^{S_i,S_i;R_i,\tau_i}_{f_i,e_i;p_i} 
B^{R_i; \,  tr}_{p_i} 
\Big]  \crcr
&&
= N^{-6n} \Big[  \prod_{i=1}^3 \delta_{R'_i R_i } \Big]\delta_{\tau'\tau}
 \sum_{S_i,\tau_i}
  \Big[ \prod_{i=1}^3   \Dim_N(S_i) \Big]
\Big[\sum_{b_i,c_i,p_i}
D^{S_i}_{b_ic_i}(\xi) 
C^{S_i,S_i;R_i,\tau_i}_{b_i,c_i;p_i}
B^{R_i; \,  tr}_{p_i}
\Big] ^2 \,. 
\eea
The evaluation finally yields
\bea
&& 
\la  O^{R_1, R_2, R_3, \tau}   \, O^{R'_1, R'_2, R'_3, \tau'}   \ra  = 
\Big[  \prod_{i=1}^3 \delta_{R'_i R_i } \Big]\delta_{\tau'\tau}
F(R_1, R_2, R_3,\tau) \crcr
&&
F(R_1, R_2, R_3,\tau) =  \sum_{S_i,\tau_i}
  \Big[ \prod_{i=1}^3   \Dim_N(S_i) \Big]
\Big[\sum_{b_i,c_i,p_i}
D^{S_i}_{b_ic_i}(\xi) 
C^{S_i,S_i;R_i,\tau_i}_{b_i,c_i;p_i}
B^{R_i; \,  tr}_{p_i}
\Big] ^2 \; .
\eea
This is \eqref{correlatorRepr} and 
 implies the orthogonality of the representation theoretic base $\{O^{R_1, R_2, R_3, \tau} \}$.

\section{Codes}
\label{app:mathsage}

We list here some algorithms which count the number of orthogonal invariants as given 
in the text. We use Mathematica and Sage softwares in the following. 

\

\noindent{\bf Mathematica code for $Z_d(t)$.}
We wish to compute the number $Z_{d}(2n)$ of rank $d$ orthogonal invariants made with 
$2n$ tensors. In order to obtain that number, we first code the generating function,
denoted {\tt Z[X, t]}, of the counting of the number of elements of  the wreath product $S_n[S_2]$  in a certain 
conjugacy class of $S_{2n}$. Doing this, we use the built-in function {\tt Count[list, pattern]} which counts the number of elements in a {\tt list} matching a {\tt pattern}.
Then, we  extract a coefficient of $t^n$ in ${\tt Z[X, t]}$ that 
is involved in ${\tt Zd[X, n, d]}$ that encodes $Z_d(2n)$. 
We finally give the counting for ranks $3$ and $4$, successively,
for $n=1,\dots, 10$.

\

{\footnotesize
\begin{verbatim}
X = Array[x, 20];
PP[n_] := IntegerPartitions[n]
Sym[q_, n_] := Product[i^(Count[q, i]) Count[q, i]!, {i, 1, n}]
Symd[X_, k_, q_] := Product[(X[[k*l]]/l)^(Count[q, l])/(Count[q, l]!), {l, 1, 2}]

Z[X_, t_] := Product[Exp[(t^i/i)*Sum[Symd[X, i, PP[2][[j]]], {j, 1, Length[PP[2]]}]], 
            {i, 1, 15}]
Zprim[X_, n_] := Coefficient[Series[Z[X, t], {t, 0, n}], t^n]
CC[X_, n_, q_] := Coefficient[Zprim[X, n], Product[X[[i]]^(Count[q, i]), {i, 1, 2*n}]]
Zd[X_, n_, d_] := Sum[(CC[X, n, PP[2*n][[i]]])^d*(Sym[PP[2*n][[i]], 2*n])^(d - 1), 
                    {i, 1, Length[PP[2*n]]}]

Table[Zd[X, i, 3], {i, 1, 10}]

(out) {1, 5, 16, 86, 448, 3580, 34981, 448628, 6854130, 121173330}

Table[Zd[X, i, 4], {i, 1, 10}] 

(out) {1, 14, 132, 4154, 234004, 24791668, 3844630928, 809199787472, 220685007519070, 
75649235368772418}
\end{verbatim}
}

\

\noindent{\bf Mathematica code: Counting with Hermite polynomials.}
This part is dedicated to the implementation of an algorithm realizing Read's enumeration 
of $k$-regular graphs on $2n$ vertices with edges of $k$ different colors where one of each color is at every vertex. We want to compare Read's results with the previous sequences. 

Read's generating function that encodes the above enumeration denotes {\tt ZR[t, d, n]},
in the following program. Then, {\tt ZR[d, n]} yields the counting at rank $d$ with $2n$ vertices 
and that is given by the coefficient  of $t^n$ in {\tt ZR[t, d, n]}. 
We evaluate $Z_3(2n)$ and $Z_4(2n)$ for the ranks $3$ and $4$, respectively, and confirm
that the results of Read match with the previous results. 

Next, the number of connected rank $d$ tensor invariants made with $2n$ tensors, written 
below {\tt ZRc[d, n]},  can be obtained using the plethystic logarithm (Plog) function. The Plog function $\plog Z_{d}(t)$, denoted {\tt Plog[ZR, t, d, n]}, is defined with the {\tt MoebiusMu} implementing the M\"obius function.

{\footnotesize
\begin{verbatim}
A[p_, v_] := (I Sqrt[p])^v HermiteH[v, 1/(2 I Sqrt[p])]
ZR[t_, d_, n_] = 1;

For[m = 0, m <= 20, m++
     {If[OddQ[m], 
    	 Phi[m, t_, d_, n_] := (Sum[((2 v)!)^(d - 1)/(v!)^(d)*(m^(d - 2)/2^d)^
         v t^(m v), {v, 0, n}]), 
    	 Phi[m, t_, d_, n_] := (Sum[(A[m/2, v])^d/(v! m^v) t^(m v/2), {v, 0, n}])]
      };
    ZR[t_, d_, n_] = ZR[t, d, n]*Phi[m, t, d, n]
]

ZR[d_, n_] := Coefficient[Series[ZR[t, d, n], {t, 0, n}], t^n]
Plog[F_, t_, d_, n_] := Sum[MoebiusMu[i]/i Log[F[t^i, d, n]], {i, 1, n}]
ZRc[d_, n_] := Coefficient[Series[Plog[ZR, t, d, n], {t, 0, n}], t^n]

Table[ZR[3, i], {i, 1, 10}]

(Out) {1, 5, 16, 86, 448, 3580, 34981, 448628, 6854130, 121173330}

Table[ZR[4, i], {i, 1, 10}]

(Out) {1, 14, 132, 4154, 234004, 24791668, 3844630928, 809199787472, 220685007519070, 
75649235368772418}

Table[ZRc[3, i], {i, 1, 10}]

(Out) {1, 4, 11, 60, 318, 2806, 29359, 396196, 6231794, 112137138}

Table[ZRc[4, i], {i, 1, 10}]

(Out) {1, 13, 118, 3931, 228316, 24499085, 3816396556, 805001547991, 219822379032704, 
75417509926065404}

\end{verbatim}
}

\noindent{\bf Sage code: Counting from the sum of Kroneckers  in rank $d=3$.}
We provide here a Sage code that recovers the same counting
through the sum of constrained Kronecker coefficients with even partitions \eqref{kroneven}. 

We need the library {\tt SymmetricFunctions(QQ)} which introduces symmetric
functions. The Kronecker coefficient associated with 
three partitions $R,S$ and $T$ deduces as the usual Hall scalar product of
Schur symmetric functions. In the following,  $s(S)$ is the Schur function associated with 
the partition $S$.

{\footnotesize

\begin{verbatim}
s = SymmetricFunctions(QQ).s()
for n in range(1,4) :
    Total=0
    for R in Partitions(2*n) :
        i=0
        rep=0
        while ( (i < R.length()) & (rep==0) ):
            if ( (R.get_part(i)%2) !=0 ):
                rep = 1
            i=i+1
        if (rep ==0) :
            for S in Partitions (2*n) :
                j=0
                rep2=0
                while ( (j < S.length()) & (rep2==0) ):
                    if ( (S.get_part(j)%2) !=0 ):
                        rep2 = 1
                    j=j+1
                if (rep2 ==0) :
                    for T in Partitions (2*n) :
                        k=0
                        rep3=0
                        while ( (k < T.length()) & (rep3==0) ):
                            if ( (T.get_part(k)%2) !=0 ):
                                rep3 = 1
                            k=k+1
                        if (rep3 ==0) :
                            a = ( s(S).itensor(s(T)) ).scalar ( s(R) )
                            Total =Total+a
 
    print "Number of invariants at 2n =", 2*n, "is", Total  

(out) Number of invariants at 2n = 2 is 1
Number of invariants at 2n = 4 is 5
Number of invariants at 2n = 6 is 16
Number of invariants at 2n = 8 is 86

\end{verbatim}
}

\noindent{\bf Sage code: The symplectic $K_4$ invariant is not vanishing at $d=3$.}
The present Sage code computes a specific invariant, given by complete graph $K_4$ with colored edges. The tensor rank is $d=3$ and the symplectic 
group $Sp(2N=4)$. We then extract a coefficient of the resulting polynomial
which does not vanish. Thus this $Sp(4)$ invariant exists.

The list {\tt T} of variables denoted {\tt T$_-$ijk}  represents the rank 3 tensor. We then  need bijections to map   {\tt T[l]} $ \leftrightarrow $ {\tt T$_-$ijk}. 
This is the work  of  {\tt f} and  {\tt f$_-$inv}.  {\tt J4} is the symplectic matrix of size $2N = 4$. 
To speed up the computation, whenever possible, we perform multiplications outside 
the cascade of internal loops when the factors multiplied do not involve the variable of that loop. 

{\footnotesize

\begin{verbatim}

T =[]
N = 4
for i in range(N): 
    for j in range(N):
        for k in range(N):
            T.append(var('T_'+str(i)+str(j)+str(k)))

J4 = [ [0, 0,  1, 0], [0, 0,  0, 1], [-1, 0, 0, 0], [0, -1, 0, 0] ]

def f(x,N):
    a,b,c = var ('a','b','c')
    a = x % N
    b = (x//N) % N
    c = (x//(N^2)) % N
    return c, b, a

def f_inv(x,y,z,N):
    return x*N^2 + y*N + z
 

N,t,A,TAB,TABB,TABC = var ('N','t','A','TAB','TABB','TABC')
TABCC,TABCD,TABCDD = var ('TABCC','TABCD','TABCDD')
t = 0 
N = 4
for a1 in range(N) :
  for a2 in range(N) : 
    for a3 in range(N) : 
      A = f_inv(a1,a2,a3,N)
      for b1 in range(N) :
        TAB = J4[a1][b1]
         for b2 in range(N) : 
           for b3 in range(N) : 
             TABB= TAB*T[A]*T[f_inv(b1,b2,b3,N)]
             for c1 in range(N) :
               for c2 in range(N) : 
                 TABC = TABB*J4[a2][c2] 
                  for c3 in range(n) :
                    TABCC = TABC*T[f_inv(c1,c2,c3,N)]*J4[b3][c3] 
                     for d1 in range(N) :
                       TABCD = TABCC*J4[c1][d1]
                        for d2 in range(N) :
                          TABCDD = TABCD*J4[b2][d2]
                           for d3 in range(N) :
                             t = t + TABCDD*T[f_inv(d1,d2,d3,N)]*J4[a3][d3] 
                                                                             

t.coefficient(T_000*T_000)
(Out) 0

t.coefficient(T_000)
(Out) 4*T_032*T_212*T_220 - 4*T_023*T_202*T_221 + 4*T_022*T_203*T_221 +
 4*T_032*T_213*T_221 + 4*T_023*T_201*T_222 - 4*T_021*T_203*T_222 - 4*T_032*T_210*T_222
- 4*T_022*T_201*T_223 + 4*T_021*T_202*T_223 - 4*T_032*T_211*T_223 - 4*T_022*T_212*T_230 
+ 4*T_012*T_222*T_230 - 4*T_023*T_212*T_231 + 4*T_012*T_223*T_231 + 4*T_022*T_210*T_232 
+ 4*T_023*T_211*T_232 - 4*T_021*T_213*T_232 - 4*T_012*T_220*T_232 + 4*T_021*T_212*T_233 
- 4*T_012*T_221*T_233 - 4*T_122*T_220*T_302 - 4*T_123*T_221*T_302 + 4*T_120*T_222*T_302
+ 4*T_121*T_223*T_302 - 4*T_122*T_230*T_312 - 4*T_123*T_231*T_312 + 4*T_120*T_232*T_312
+ 4*T_121*T_233*T_312 + 4*T_122*T_202*T_320 + 4*T_132*T_212*T_320 - 4*T_102*T_222*T_320
- 4*T_112*T_232*T_320 + 4*T_122*T_203*T_321 + 4*T_132*T_213*T_321 - 4*T_102*T_223*T_321
- 4*T_112*T_233*T_321 + 4*T_123*T_201*T_322 - 4*T_120*T_202*T_322 - 4*T_121*T_203*T_322
- 4*T_132*T_210*T_322 + 4*T_102*T_220*T_322 + 4*T_112*T_230*T_322 - 4*T_122*T_201*T_323
- 4*T_132*T_211*T_323 + 4*T_102*T_221*T_323 + 4*T_112*T_231*T_323 + 4*T_122*T_210*T_332 
+ 4*T_123*T_211*T_332 - 4*T_120*T_212*T_332 - 4*T_121*T_213*T_332
          
t.coefficient(T_032*T_212*T_220)
(Out)  4*T_000

\end{verbatim}
}

\end{document}